\documentclass[12pt]{article}
\usepackage{amsfonts}
\usepackage{amsmath}
\usepackage{amssymb}
\usepackage{amsthm}
\usepackage{bbm}

\usepackage[utf8]{inputenc}
\usepackage[margin=2cm]{geometry}
\usepackage{geometry}
\usepackage{hyperref}
\usepackage{graphicx}
\usepackage{amsmath}
\usepackage{cite}
\usepackage{hyperref}
\usepackage{tensor}
\usepackage{cases}
\usepackage{appendix}
\usepackage{multirow}
\usepackage{booktabs}
\usepackage{color}
\usepackage[font=small]{subfig}
\usepackage[skip=8pt,labelfont={bf,sc},textfont=small]{caption}

\def\[#1\]{\begin{align}#1\end{align}}

\def \nn {\nonumber}

\def \dd{\mathrm{d}}
\def \D {\Delta}
\def \e {\epsilon}

\def \ra{\rangle}
\def \la{\langle}

\def \mo{\mathcal{O}}

\def \tr{\text{tr}}

\begin{document}
\begin{titlepage}
\vspace{0.5cm}
\begin{center}
{\Large \bf {{Relative R\'enyi Entropy} Under Local Quenches in 2D CFTs}}
\lineskip .75em
\vskip 2.5cm
Zi-Xuan Zhao$^{a}$ %The unusual ordering of authors instead of the standard alphabetical one in the hep-th community is to ensure that the student receives proper recognition for the student's contribution under the current practice in Jilin University.}
, Song He$^{b,c,d}$\footnote{hesong@nbu.edu.cn; Corresponding author},
Hao Ouyang$^{a,}$\footnote{haoouyang@jlu.edu.cn; Corresponding author}, Hong-an Zeng$^{a,}$\footnote{zengha20@mails.jlu.edu.cn; Corresponding author},
Yu-Xuan Zhang$^{e,}$\footnote{zhangyuxuan@ucas.ac.cn; Corresponding author}

\vskip 2.5em
{\normalsize\it
$^{a}$Center for Theoretical Physics and College of Physics, Jilin University,\\ Changchun 130012, People's Republic of China\\
$^{b}$Institute of Fundamental Physics and Quantum Technology,
Ningbo University, \\Ningbo, 315211, People's Republic of China\\
$^{c}$School of Physical Science and Technology, Ningbo University,\\ Ningbo, 315211, People's Republic of China\\
$^{d}$Max Planck Institute for Gravitational Physics (Albert Einstein Institute),\\
Am M\"uhlenberg 1, 14476 Golm, Germany\\
$^{e}$Kavli Institute for Theoretical Sciences, University of Chinese Academy of Sciences,\\ Beijing
100190, People's Republic of China}\\
\vskip 1.0em
\vskip 3.0em
\end{center}
\begin{abstract}
%We studied the relative Rényi entropy (RRE) under different local quenches in rational CFTs (RCFTs) and holographic CFTs. In RCFTs, the RRE between reduced-density matrices excited by different operators evolves over time as a monotonic function that depends on certain finite-dimensional matrices. In some cases, the RRE is symmetric for two reduced-density matrices, so we investigated its relationship with the trace squared distance. However, relative entropy sometimes fails to distinguish seemingly different operators, primarily because it can only differentiate the information that has entered the subsystem regarding the reduced density matrices. In holographic CFTs, a special analytic continuation of the RRE can provide a pathway to obtaining the geometric structure of the entanglement wedge, offering a broader perspective for understanding bulk geometry.

We study the relative R\'enyi entropy (RRE) under local quenches in two-dimensional conformal field theories (CFTs), focusing on rational CFTs (RCFTs) and holographic CFTs. In RCFTs, the RRE evolves as a monotonic function over time, depending on finite-dimensional matrices. It is sometimes symmetric, prompting an investigation into its relationship with the squared trace distance. We also observe that relative entropy can fail to distinguish between operators, as it only captures information entering/exiting the subsystem. In holographic CFTs, an analytic continuation of the RRE reveals insights into the entanglement wedge, offering a new perspective on bulk geometry in AdS/CFT. Our results deepen the understanding of quantum information measures in RCFTs and holographic CFTs, highlighting connections to distinguishability and bulk reconstruction.

\end{abstract}
\end{titlepage}
\baselineskip=0.7cm
\tableofcontents

\section{Introduction}
The AdS/CFT correspondence \cite{Maldacena:1997re, Gubser:1998bc, Witten:1998qj} provides a powerful holographic framework linking quantum gravity in an asymptotically anti-de Sitter (AdS) space to a CFT on its boundary. This correspondence has allowed quantum information theory to contribute significantly to our understanding of gauge/gravity duality and quantum gravity, leading to significant advances in high-energy physics, particularly in the areas of quantum entanglement \cite{Casini:2004bw, Calabrese:2004eu, Kitaev:2005dm, Casini:2016fgb, Nishioka:2018khk, Witten:2018zxz, Casini:2022rlv}, the emergence of spacetime geometry \cite{VanRaamsdonk:2010pw, Maldacena:2013xja, Rangamani:2016dms}, and the black hole information paradox \cite{Hawking:1976ra, Mathur:2009hf, Almheiri:2012rt, Penington:2019npb, Almheiri:2019psf}. A prominent example is the Ryu-Takayanagi (RT) formula \cite{Ryu:2006bv, Hubeny:2007xt, Chen:2019lcd}, which relates the entanglement entropy of the boundary quantum field theory (QFT) to the area of a codimension-2 minimal surface in the dual spacetime. This formula has been extended to higher-order gravity theories \cite{Dong:2013qoa, Camps:2013zua, Miao:2014nxa} and scenarios with quantum corrections \cite{Faulkner:2013ana, Engelhardt:2014gca}.

%\textcolor{red}{Overlap with 2305.10984:}The AdS/CFT correspondence\cite{Maldacena:1997re, Gubser:1998bc, Witten:1998qj} provides a holographic description of a quantum gravity theory in an asymptotically anti-de Sitter (AdS) spacetime with a conformal field theory (CFT) at its asymptotic boundary. Its discovery has inspired significant exploration regarding quantum information theory in the high-energy physics community in recent years, covering a range of topics such as quantum entanglement\cite{Casini:2004bw, Calabrese:2004eu, Kitaev:2005dm, Casini:2016fgb, Nishioka:2018khk, Witten:2018zxz, Casini:2022rlv}, the emergence of geometry \cite{VanRaamsdonk:2010pw, Maldacena:2013xja, Rangamani:2016dms}, and the black hole information paradox \cite{Hawking:1976ra, Mathur:2009hf, Almheiri:2012rt, Penington:2019npb, Almheiri:2019psf}.\textcolor{red}{Overlap with 2102.01898:} As a result, the quantum information theoretic considerations have provided various viewpoints in the study of gauge/gravity duality and quantum gravity. One of the famous examples is the Ryu-Takayanagi (RT) formula \cite{Ryu:2006bv,Hubeny:2007xt,Chen:2019lcd}, which connects the area of a codimension-2 minimal surface in the dual spacetime and the entanglement entropy of the boundary QFT. and has been generalized into higher order gravity theory \cite{Dong:2013qoa,Camps:2013zua,Miao:2014nxa} and cases with quantum corrections \cite{Faulkner:2013ana,Engelhardt:2014gca}.

Most previous studies have focused on the entanglement of a subsystem within a given quantum state, for instance, the dynamics of subsystem entanglement entropy for a non-equilibrium state created by local quenches \cite{Alcaraz:2011tn,Bhattacharya:2012mi, Nozaki:2014uaa, Nozaki:2014hna, He:2014mwa, Caputa:2014vaa, Caputa:2014eta, Asplund:2014coa, Caputa:2015waa, Caputa:2015tua, Chen:2015usa, Nozaki:2015mca, Caputa:2015qbk, Nozaki:2016mcy, Numasawa:2016kmo, Caputa:2017tju, Nozaki:2017hby, Kusuki:2017upd, Kusuki:2018wpa, Zhang:2019kwu, Kusuki:2019gjs, Kusuki:2019evw, Kusuki:2019avm,Kudler-Flam:2020xqu,Agon:2020fqs,Nandy:2021hmk,Kawamoto:2022etl,Gyongyosi:2024muh,Mao:2024cnm,Mao:2025hkp} or  global quenches \cite{Calabrese:2005in, Hartman:2013qma, Asplund:2015eha, Calabrese:2016xau, Caputa:2017ixa, Nishida:2017hqd, Wen:2018vux, MacCormack:2018rwq, Wen:2018agb, Fujita:2018lfj, Kim:2019egi, Kudler-Flam:2020url, Wen:2020wee, Han:2020kwp, Goto:2021gve, Das:2021gts, Goto:2023wai, Kudler-Flam:2023ahk, Nozaki:2023fkx, Miyata:2024gvr, Bai:2024azk, Das:2024lra, Jiang:2024hgt, Bernamonti:2024fgx} protocols. 
A natural extension is exploring whether other quantum information concepts can offer further insights when comparing two quantum states defined on the same Hilbert space. In this context, a particularly interesting quantity is the so-called relative entropy, which, for two given (reduced) density matrices $\rho$ and $\sigma$, is defined as:
%\textcolor{red}{Overlap with 1612.00659:} So far, the large majority of these studies focused on the entanglement of a subsystem of a given quantum state \cite{He:2014mwa,Chen:2015usa,Mollabashi:2020yie,He:2023eap}. It is a very natural question whether, more generally speaking, exploring other quantum information concepts could provide more insights when considering two different quantum states (obviously defined on the same Hilbert space). In this respect, an interesting quantity to look at is the so-called relative entropy, relative entropy that for two given (reduced) density matrices $\rho$ and $\sigma$, is defined as
\[
S(\rho||\sigma)=\tr[\rho\log\rho]-\tr[\rho\log\sigma]\label{eq1.1}.
\]
This quantity can be interpreted as a measure of the distinguishability between quantum states, serving as an asymmetric ``distance" between $\rho$ and $\sigma$. Although not an entanglement measure, relative entropy is closely related to various entanglement measures \cite{Vedral:2002zz}.
%which can be interpreted as a measure of distinguishability of quantum states, being a sort of (asymmetric) “distance” between $\rho$ and $\sigma$. It is not an entanglement measure itself, but nonetheless has a connection with several entanglement measures \cite{Vedral:2002zz}.

Unlike entanglement entropy, which suffers from ultraviolet divergences in quantum field theory, relative entropy is finite and well-defined \cite{Witten:2018zxz}, making it a central focus of many studies \cite{Casini:2016udt, Casini:2008cr, Blanco:2013joa, Jafferis:2015del, Lashkari:2014yva, Lashkari:2015dia, Sarosi:2016oks, Ugajin:2016opf, Sarosi:2016atx, Balasubramanian:2014bfa, Ruggiero:2016khg, Ugajin:2018rwd, Bao:2019aol, Brehm:2020zri,deBoer:2023lrd, Ugajin:2020dyd}. It is closely connected to the modular Hamiltonian \cite{Lashkari:2015dia} and provides valuable insights into various areas, including condensed matter systems \cite{Zhou:2008zza, Zanardi:2006mgo}, and the Bekenstein bound \cite{Casini:2008cr}, as well as its holographic counterpart \cite{Blanco:2013joa, Jafferis:2015del, Bao:2019aol}.

%Contrarily to the entanglement entropy which in a quantum field theory framework suffers from the problem of ultraviolet divergences, the relative entropy is finite and therefore well defined also in field theory and has already taken a central role given the number of papers devoted to it, see e.g. \cite{Casini:2016udt,Casini:2008cr,Blanco:2013joa,Jafferis:2015del,Lashkari:2014yva,Lashkari:2015dia,Sarosi:2016oks,Ugajin:2016opf,Sarosi:2016atx,Balasubramanian:2014bfa,Ruggiero:2016khg,Ugajin:2018rwd,Bao:2019aol}.It is related to the modular Hamiltonian \cite{Lashkari:2015dia}, and may also give useful insights also in the study of condensed matter systems \cite{Zhou:2008zza,Zanardi:2006mgo}. It has also been considered in connection to the laws of black hole thermodynamics and the Bekenstein bound \cite{Casini:2008cr}. Its holographic version has been discussed as well \cite{Blanco:2013joa,Jafferis:2015del,Bao:2019aol}.
In QFT, relative entropy is derived by modifying the replica trick for entanglement entropy \cite{Calabrese:2004eu}, as introduced by Lashkari \cite{Lashkari:2014yva}. This method involves defining { the so-called sandwiched relative R\'enyi entropy \cite{Wilde:2013bdg,Muller-Lennert:2013liu}
\[
S_\alpha(\rho||\sigma)=\begin{cases}
    \frac{1}{\alpha-1}\log \tr([\sigma^{\frac{1-\alpha}{2\alpha}}\rho\sigma^{\frac{1-\alpha}{2\alpha}}]^\alpha) \quad \text{if} \rho\not\perp\sigma\wedge(\sigma\gg\rho\vee\alpha<1)\\
    \infty \quad \text{else}
\end{cases}
\]
for any $\alpha\in(1/2,1)\cup(1,\infty)$. For the range $\alpha\in(0,1/2)$, it is more natural to use the Petz $\alpha$-relative entropy \cite{Petz:1986naj} defined as 
\[
S_\alpha(\rho||\sigma)=\frac{1}{\alpha-1}\log \tr[\rho^\alpha\sigma^{1-\alpha}].
\]
An alternative approach \cite{Lashkari:2015dia} used for computing the matrix elements of the modular Hamiltonian corresponding to excited states involves introducing a Rényi-like one-parameter quantity $\mathrm{Tr}(\rho\sigma^{k-1})$, which for integer $k$ serves as a generalized partition function or correlation function on an $k$-sheeted Riemann surface, thereby breaking the $Z_k$ symmetry among replicas. The relative entropy is then obtained through the replica limit $k\to 1$ of the RRE\footnote{For convenience, we refer to the following Rényi-like one-parameter quantity as RRE throughout this work.} defined as
%In a quantum field theory, the relative entropy can be obtained by a variation of the replica trick for the entanglement entropy \cite{Calabrese:2004eu} which has been introduced by Lashkari \cite{Lashkari:2014yva} and later refined by the  same author \cite{Lashkari:2015dia}. The main idea is to introduce the new quantity $\mathrm{Tr}(\rho\sigma^n)$ that for $n$ integer is a generalized partition function or correlation function on a $n$-sheeted Riemann surface which breaks the $Z_n$ symmetry among replicas. The relative entropy is given by the following replica limit \cite{Lashkari:2015dia}
\[
S_k(\rho||\sigma)=\frac{1}{k-1}\left(\log\tr[\rho^k]-\log\tr[\rho\sigma^{k-1}]\right),
\]
provided that the parameter $k$ can be analytically continued from integer to complex values. This general method, at least in principle, allows the computation of relative entropy in any quantum field theory.  However, to date, only a few direct calculations of relative entropy have been performed in 1+1-dimensional CFTs \cite{Lashkari:2014yva, Lashkari:2015dia, Ugajin:2016opf, Sarosi:2016oks, Ruggiero:2016khg}.

While the RRE does not require certain formal properties of the sandwiched RRE—such as UV finiteness, non-negativity, and satisfaction of the data processing inequality \cite{Wilde:2013bdg}\footnote{We sincerely thank the anonymous referee for highlighting this important point.}— and it is still unknown whether this quantity has a quantum information interpretation, it remains worthy of investigation because it provides a more intuitive realization of the replica method while circumventing the subtle analytical continuation required when extending from positive integer to fractional values. Our preference for RRE in this work does not imply its superiority over sandwiched RRE or Petz $\alpha$-RRE. The RRE was chosen purely for its clearer replica geometry visualization and more straightforward computational implementation in our framework. While one may alternatively adopt sandwiched RRE or Petz $\alpha$-RRE formulations, our computational framework demonstrates they would yield equivalent results. }
%whenever the analytic continuation of the parameter $n$ from integer to complex values is obtainable. This method is completely general and permits (at least in principle) the computation of the relative entropy in a generic quantum field theory. However, up to now, only a few direct calculations of relative entropy have been performed in 1+1 dimensional CFT \cite{Lashkari:2014yva,Lashkari:2015dia,Ugajin:2016opf,Sarosi:2016oks,Ruggiero:2016khg} and only very recently have some results for arbitrary dimensions appeared \cite{Sarosi:2016atx}.

{This paper will investigate the RRE in RCFTs and holographic CFTs under local quenches.  Although the information-theoretic properties of conformal descendants have been explored previously \cite{Palmai:2014jqa, Caputa:2015tua, Taddia:2016dbm, Brehm:2020zri, Chen:2015usa, He:2023eap}, it has been found that operators in the same conformal family encode the same information in the context of entanglement entropy \cite{Caputa:2015tua, Chen:2015usa} or pseudo entropy \cite{Guo:2022sfl, He:2023eap}, typically characterized by the quantum dimension $d_{\mo}$ of the corresponding primary operator $\mo$.}{ In \cite{Caputa:2015tua}, the authors pointed out that when considering the Schmidt decomposition 
of quantum states, different descendant states may share the same Schmidt coefficients 
while possessing different Schmidt bases, leading to identical entanglement entropy. This information, arising from the change in the Schmidt basis, can be captured by the RRE. 
} Therefore, investigating the RRE under different local (descendant) quenches in RCFTs may deepen our understanding of the role of quantum information in CFTs by evaluating how distinguishable different quantum states are. Additionally, examining the information properties of conformal descendant operators in RCFTs through RRE may reveal their unique quantum information characteristics, aiding in distinguishing and understanding the information carried by different operators within the same conformal family. RRE may closely relate to geometric structures in bulk in the context of holographic duality since its special case, fidelity, can reconstruct the entanglement wedge \cite{Suzuki:2019xdq}. Investigating RRE under local quenches may allow us to explore how quantum information measures can reconstruct bulk geometry, offering new perspectives and tools for the AdS/CFT correspondence.

The rest of this paper is organized as follows. In section \ref{section2}, we briefly review the replica method for locally excited states in 2D CFTs and provide our convention and some useful formulae for the later calculations. In section \ref{section3}, we mainly focus on the RRE of locally descendant excited states in RCFTs. For simplicity, we study the cases in which finite holomorphic Virasoro generators generate the descendants. More general and complicated situations are discussed in section \ref{section4}, where we derive the full-time evolution of the $k^{\rm th}$ RRE for the generic combination states. In section~\ref{section5}, we investigate RRE under local quenches in holographic CFTs. We end with conclusions and discussions in section \ref{section6}. Some calculation details and codes are presented in the appendices.
{\section{Setup in 2D CFTs}\label{section2}}
{Our focus is on the RRE between two time-evolved density matrices $\rho$ and $\sigma$, generated by two different operators:
\[
\rho=\frac{e^{-iHt}e^{-\epsilon H}|\psi\ra\la\psi|e^{-\epsilon H}e^{iHt}}{\la\psi|e^{-2\epsilon H}|\psi\ra},\quad
\sigma=\frac{e^{-iHt'}e^{-\epsilon H}|\psi'\ra\la\psi'|e^{-\epsilon H}e^{iHt'}}{\la\psi'|e^{-2\epsilon H}|\psi'\ra},\label{eq2.1}
\]
where $|\psi\rangle = \mathcal{O}(x)|\Omega\rangle$ and $|\psi'\rangle = \mathcal{O}'(x')|\Omega\rangle$. Note that an infinitesimal positive parameter $\epsilon$ has been introduced to suppress high-energy modes \cite{Calabrese:2005in}.}

{By tracing out the degrees of freedom of $A^c$ (the complement of the subsystem $A$), we can obtain two reduced density matrices $\rho_{A}(t)=\tr_{A^c}[\rho(t)]$ and $\sigma_{A}(t)=\tr_{A^c}[\sigma(t)]$. 
The $k^{\rm th}$ RRE of $\rho_A$ and $\sigma_A$ can be derived through the replica method
\[
S_{k}(\rho_A||\sigma_A)=\frac{1}{k-1}&\left(\log\tr [\rho_A^k]-\log\tr[\rho\sigma_A^{k-1}]\right)\nn\\
=\frac{1}{k-1}&\left(\log \frac{\la\mo(w_1,\bar w_1)\dots\mo(w_{2k},\bar w_{2k})\ra_{\Sigma_k}}{\la\mo(w_1,\bar w_1)\mo(w_2,\bar w_1)\ra_{\Sigma_1}^k}\right .\nn\\
-&\left.\log \frac{\la \mo(w_1,\bar w_1)\mo(w_2,\bar w_2)\mo'(w'_3,\bar w'_3)\dots\mo'(w'_{2k},\bar w'_{2k})\ra_{\Sigma_k}}{\la \mo(w_1,\bar w_1)\mo(w_2,\bar w_2)\ra_{\Sigma_1} \la\mo'(w'_1,\bar w'_1)\mo'(w'_2,\bar w'_2)\ra_{\Sigma_1}^{k-1}}\right).\label{eq2.2}
\]
In \eqref{eq2.2}, $\Sigma_k$ denotes a $k$-sheeted Riemann surface with cuts on each copy corresponding to $A$, and $(w_{2j-1},\bar w_{2j-1})$ and $(w_{2j},\bar w_{2j})$ are coordinates on the $j$-th sheet surface. Although there are Hermitian conjugates (daggers) when changing a bra to a ket in Eq.~\eqref{eq2.2}, we omit them for clarity in this paper. One of the standard approaches to computing $2k$-point functions on $\Sigma_k$ is to transform them into the $2k$-point functions on $\Sigma_1$ (i.e., the complex plane) by utilizing a certain conformal mapping \cite{He:2014mwa, Chen:2015usa}, which we summarize in the Appendix \ref{app111}, along with more details about the Wick rotation of the Euclidean time.}

\section{Relative R\'enyi entropy for different states}\label{section3}
In RCFTs, it is known that the excess of R\'enyi entropy for primary and descendant operators \cite{He:2014mwa, Caputa:2015tua, Chen:2015usa} and the pseudo-R\'enyi entropy for primary and descendant operators \cite{Guo:2022sfl, He:2023eap} both saturate to a constant value. This constant is equal to the logarithm of the quantum dimension of the associated primary operator. {Specifically, if we consider two density matrices $\rho$ and $\sigma$ constructed from a primary operator $\mathcal{O}$ and a descendant operator $L_{-n}\mathcal{O}$, respectively, i.e.
	\[
	\rho(t)=\frac{e^{-(\epsilon+it)H}\mo(x)|\Omega\ra\la\Omega|\mo(x)e^{-(\epsilon-it) H}}{\la\Omega|\mo(x)e^{-2\epsilon H}\mo(x)|\Omega\ra},\quad \sigma(t)=\frac{e^{-(\epsilon+it)H} L_{-n}\mo(x)|\Omega\ra\la\Omega|L_{-n}\mo(x)e^{-(\epsilon-it) H}}{\la\Omega| L_{-n}\mo(x)e^{-2\epsilon H}L_{-n}\mo(x)|\Omega\ra},\quad 
	\]
	 the subsystem entanglement  entropies of $\rho$ and $\sigma$ are identical. For example, taking a semi-infinte subsystem $A=[0,\infty]$, we have 
	 \[
S(\rho_A(t))=S(\sigma_A(t))=\begin{cases} 
	 	0 &  t<|x| \\
	 	\log d_{\mo} & t>|x|.
	 \end{cases}\label{vacuum_primary}
	 \] This result implies that primary and descendant operators are indistinguishable regarding entanglement entropy. Moreover, if we consider a transition matrix $\mathcal{T}^{\mathcal{O}|L_{-n}\mathcal{O}}:=\frac{\mo|\Omega\ra\la\Omega|L_{-n}\mo}{\la L_{-n}\mo|\mathcal{O}\ra}$, we can find the pseudo-entropy \cite{Nakata:2020luh} of a semi-infinte subsystem $A$, defined as $S(\mathcal{T}_A^{\mathcal{O}|L_{-n}\mathcal{O}}):=-\tr \big[\mathcal{T}_A^{\mathcal{O}|L_{-n}\mathcal{O}}\log \mathcal{T}_A^{\mathcal{O}|L_{-n}\mathcal{O}}\big]$ (with $T_A^{\mathcal{O}|L_{-n}\mathcal{O}}:=\tr_{A^c}\big[T^{\mathcal{O}|L_{-n}\mathcal{O}}\big]$) also saturate to $\log d_{\mo}$ at late times \cite{He:2023eap}, which means all descendant information vanish.} %In other words, these quantities are defined on an equivalent class.} 
     Relative entropy, proposed as a measure of the distinguishability between two states, plays a central role in quantum information theory \cite{Lashkari:2014yva}. { To under this ``fine structure'' of conformal family,} this section will examine RRE for different states excited at the same position within the same conformal family in RCFTs to investigate how information-theoretic quantities of descendants might be discerned. For simplicity, we will choose an interval on the right of the inserted operator, noting that a conformal transformation allows us to obtain any desired insertion position.

{\subsection{$k^{\rm th}$ relative R\'enyi entropy for $|\Omega\rangle$ and $\mo(x)|\Omega\rangle$}
Let us initially explore the most straightforward case that one density matrix is constructed from a vacuum state where we set $\mo'$ equal to the identity operator $\mathbbm{1}$ while the other is built from a primary operator $\mo$ with equal chiral and antichiral conformal weights $h=\bar h=\Delta$, i.e.
\[
\rho=\frac{e^{-iHt}\mo(x,-\epsilon)|\Omega\ra\la\Omega|\mo(x,\epsilon)e^{iHt}}{\la\mo(x,\epsilon)\mo(x,-\epsilon)\ra},\quad \quad\sigma=\frac{|\Omega\ra\la\Omega|}{\la\Omega|\Omega\ra}.
\]}
{According to \eqref{eq2.2}, the $k^{\rm th}$ RRE of primary operator and identity operator is
\[
S_{k}(\rho||\sigma)=\frac{1}{k-1}\log \frac{\la\mo(w_1,\bar w_1)\dots\mo(w_{2k},\bar w_{2k})\ra_{\Sigma_k}}{\la\mo(w_1,\bar w_1)\mo(w_2,\bar w_1)\ra_{\Sigma_1}^k}-\frac{1}{k-1}\log\frac{\la \mo(w_1,\bar w_1)\mo(w_2,\bar w_2)\mathbbm{1}\dots\mathbbm{1}\ra_{\Sigma_k}}{\la\mo(w_1,\bar w_1)\mo(w_2,\bar w_1)\ra_{\Sigma_1}}.\label{3.1.1}
%=&\frac{(2\epsilon)^{-4\Delta}}{(2\epsilon)^{-2\Delta}(2k\sin \frac{\pi}{k}(t+x))^{-2\Delta}}
\]
The first term on the RHS of \eqref{3.1.1} is the increase of R\'enyi entropy of $\mo$, and its evolution has been discussed exclusively in \cite{He:2014mwa}, which would equal to $0$ at early times ($t<\left|x\right|$) while equal to $\log d_{\mo}$, where $d_{\mo}$ is the quantum dimension of $\mo$, at late times ($t>\left|x\right|$). The second term on the RHS of \eqref{3.1.1}, by utilizing the conformal map \eqref{conformal_map}, can be reformulated as 
\[
\tr(\rho \sigma^{k-1})=\frac{\left|\frac{d w_1}{d z_1}\right|^{-2\Delta}\left|\frac{d w_2}{d z_2}\right|^{-2\Delta}\la \mo(z_1,\bar z_1)\mo(z_2,\bar z_2)\mathbbm{1}\dots\mathbbm{1}\ra_{\Sigma_1}}{\la\mo(w_1,\bar w_1)\mo(w_2,\bar w_1)\ra_{\Sigma_1}}\label{3.1.2},
\]
It should be noted that Eq.~\eqref{3.1.2} is supposed to have some contributions from the vacuum. However, since this part cancels out with the vacuum contribution in the entanglement entropy \eqref{eq2.2}, \eqref{3.1.2} is just a convenient notation.}

{The early-time and late-time analysis of \eqref{3.1.2} can be found in Appendix \ref{appb1ooo}, which, combined with the growth of R\'enyi entropy of $\mo$ mentioned before \cite{He:2014mwa}, yields the following expression for the full-time evolution of the RRE
\[
S_{k}(\rho||\sigma)=\begin{cases} 
    0 &  t<|x| \\
    \frac{2\Delta}{k-1}\log \frac{k\sin \frac{\pi }{k}(t+x)}{\epsilon}-\log d_{\mo} & t>|x|.
\end{cases}\label{vacuum_primary}
\]
As time evolves, the $k^{\rm th}$ RRE of the primary operator and identity operator diverges when we take the regulator $\epsilon\to 0$.  The reason for the divergence of RRE is that the regularization we take in \eqref{eq2.1} is no longer valid since the vacuum state does not have high-energy modes to suppress. }%\textcolor{red}{In other words, $\sigma$ constructed from the vacuum state is not regularized while $\rho$ is regularized, so when we consider RRE with this local quench, it always diverges. } 

\subsection{$k^{\rm th}$ relative R\'enyi entropy for $L_{-n}\mo(x)|\Omega\rangle$ and $\mo(x)|\Omega\rangle$}
In this subsection, we will study the temporal evolution of the RRE in more complex scenarios. The two density matrices are constructed from a primary operator and a descendant operator belonging to the same conformal family, i.e.
\[
\rho=\frac{e^{-iHt}\mo(x,-\epsilon)|\Omega\ra\la\Omega|\mo(x,\epsilon)e^{iHt}}{\la\mo(x,\epsilon)\mo(x,-\epsilon)\ra},\quad \sigma=\frac{e^{-iHt}L_{-n}\mo(x,-\epsilon)|\Omega\ra\la\Omega|L_{-n}\mo(x,\epsilon)e^{iHt}}{\la L_{-n}\mo(x,\epsilon)L_{-n}\mo(x,-\epsilon)\ra}.
\]
Similar to \eqref{3.1.2}, the expression of $\tr(\rho \sigma^{k-1})$ within the definition of RRE \eqref{eq2.2} in this case is 
\[
\tr(\rho \sigma^{k-1})=\frac{\la\mo(w_1,\bar{w}_1)\mo(w_2,\bar{w}_2)L_{-n}\mo(w_3,\bar{w}_3)\dots L_{-n}\mo(w_{2k},\bar{w}_{2k})\ra_{\Sigma_k}}{\la\mo(w_1,\bar{w}_1)\mo(w_2,\bar{w}_2)\ra_{\Sigma_1}\la L_{-n}\mo(w_1,\bar{w}_1)L_{-n}\mo(w_2,\bar{w}_2)\ra_{\Sigma_1}^{k-1}}.\label{3.2.1}
%\underbrace{\dots}_{k-4 \text{\quad operators}}
\]
 The $2k$-point function above can be computed analogously to the $2k$-point functions encountered when studying Rényi entanglement entropy evolution under local operator quenches in RCFTs \cite{He:2014mwa, Chen:2015usa}. Further technical details of it are presented in Appendix \ref{appsb1} for the interested reader. In terms of the analysis in Appendix \ref{appsb1}, the full-time evolution of the RRE for a primary state and a descendant state is given by
\[
S_{k}(\rho||\sigma)=\begin{cases} 
    0 &  t<|x| \\
     \frac{1}{k-1}\log \frac{12(n+1)^2 \Delta^2}{\left(\frac{\Gamma (2n) \left(c n^2 \left(n^2-1\right)^2  +24 \Delta  (2n) (2n+1) ( n^2-1)\right)}{\Gamma (n+2) \Gamma (n+2)}+12 \Delta  (\Delta   (n+1)^2+2)\right)}& t>|x|.
\end{cases} \label{s3fulltime3.2}
\]
%\textcolor{red}{Where $c$ is the central charge which has not been defined previously, Please check!}{\color{blue} ZX: C has mentioned in Eq.(18)}
Eq.~\eqref{s3fulltime3.2} shows that although generally speaking, the RRE is asymmetrical with respect to the two density matrices that constitute it, {RRE indeed can serve as a ``measure" of the distinguishability\footnote{We emphasize that the distinguishability ``measure" discussed throughout this work does not constitute a true measure in the mathematical or quantum-informatic sense. Rather, it serves as a diagnostic indicator for different operators within the same conformal family or a probe for ``fine structures" beyond entanglement entropy.} of two states that are indistinguishable in the context of entanglement entropy.}
\subsection{$k^{\rm th}$ relative R\'enyi entropy for two general descendant states}\label{subsection3.3}
We next consider the  RRE for two general descendant states, which are generated from acting on the vacuum with two different descendant operators belonging to the same conformal family. Let the two density matrices $\rho$ and $\sigma$ take the form of
\[
\rho=&\frac{e^{-iHt}L_{-\{K_i\}}\bar{L}_{-\{\bar{K}_i\}}\mo(x,-\epsilon)|\Omega\ra\la\Omega|L_{-\{K_i\}}\bar{L}_{-\{\bar{K}_i\}}\mo(x,\epsilon)e^{iHt}}{\la L_{-\{K_i\}}\bar{L}_{-\{\bar{K}_i\}}\mo(x,\epsilon)L_{-\{K_i\}}\bar{L}_{-\{\bar{K}_i\}}\mo(x,-\epsilon)\ra},\nn\\
\sigma=&\frac{e^{-iHt}L_{-\{K'_j\}}\bar{L}_{-\{\bar{K}'_j\}}\mo(x,-\epsilon)|\Omega\ra\la\Omega|L_{-\{K'_j\}}\bar{L}_{-\{\bar{K}'_j\}}\mo(x,\epsilon)e^{iHt}}{\la L_{-\{K'_j\}}\bar{L}_{-\{\bar{K}'_j\}}\mo(x,\epsilon)L_{-\{K'_j\}}\bar{L}_{-\{\bar{K}'_j\}}\mo(x,-\epsilon)\ra},\label{quench3.3}
\]
where $L_{-\{K_i\}}\equiv L_{-k_{i_1}}L_{-k_{i_2}}...L_{-k_{i_{n_i}}}$, $(0\leq k_{i_1}\leq k_{i_2}\leq...\leq k_{i_{n_i}})$, and $L_{-\{\bar K_i\}}\equiv L_{-\bar k_{\bar i_1}}L_{-\bar k_{\bar i_2}}...L_{-\bar k_{\bar i_{n_i}}}$, $(0\leq \bar k_{\bar i_1}\leq \bar k_{\bar i_2}\leq...\leq \bar k_{\bar i_{n_i}})$ are shorthands of the compositions of Viarasoro generators (Likewise for $L_{-\{K'_j\}}$ and $L_{-\{\bar{K}'_j\}}$). The expression of $\tr(\rho \sigma^{k-1})$ now in this case is
\[
&\tr(\rho \sigma^{k-1})\nn\\
=&\frac{\la L_{-\{K_i\}}\bar{L}_{-\{\bar{K}_i\}}\mo(1)L_{-\{K_i\}}\bar{L}_{-\{\bar{K}_i\}}\mo(2)L_{-\{K'_j\}}\bar{L}_{-\{\bar{K}'_j\}}\mo(3)\dots L_{-\{K'_j\}}\bar{L}_{-\{\bar{K}'_j\}}\mo(2k)\ra_{\Sigma_k}}{\la L_{-\{K_i\}}\bar{L}_{-\{\bar{K}_i\}}\mo(1)L_{-\{K_i\}}\bar{L}_{-\{\bar{K}_i\}}\mo(2)\ra_{\Sigma_1}\la L_{-\{K'_j\}}\bar{L}_{-\{\bar{K}'_j\}}\mo(1)L_{-\{K'_j\}}\bar{L}_{-\{\bar{K}'_j\}}\mo(2)\ra_{\Sigma_1}^{k-1}},\label{s33.3.1}
\]
with $\mathcal{O}(i)\equiv\mathcal{O} (w_i, \bar w_i)$. Again, the $2k$-point function in Eq. \eqref{s33.3.1} can be evaluated as in the previous section, with details given in Appendix \ref{appa2}. According to the analysis in Appendix \ref{appa2}, the full-time evolution of $k^{\rm th}$ RRE\footnote{By employing the computational procedure developed in this subsection, one can also obtain sandwiched RRE and Petz $\alpha$-RRE, yielding the results\[
S_{sandwiched}=\begin{cases} 
    0 &  t<|x| \\
     \frac{\alpha}{\alpha-1}\log\frac{c_0\left({\{K_i\},\{K'_j\}}\right)c_0\left({\{K_i\},\{K'_j\}}\right)}{c_0\left({\{K_i\},\{K_i\}}\right)c_0\left({\{K'_j\},\{K'_j\}}\right)}& t>|x|.
\end{cases}\nn\\
S_{Petz}=\begin{cases} 
    0 &  t<|x| \\
     \frac{1}{\alpha-1}\log\frac{c_0\left({\{K_i\},\{K'_j\}}\right)c_0\left({\{K_i\},\{K'_j\}}\right)}{c_0\left({\{K_i\},\{K_i\}}\right)c_0\left({\{K'_j\},\{K'_j\}}\right)}& t>|x|.
\end{cases}
\]This indicates that starting from any form of RRE leads to equivalent results.} of $L_{-\{K_i\}}\bar{L}_{-\{\bar{K}_i\}}\mo$ and $L_{-\{K'_j\}}\bar{L}_{-\{\bar{K}'_j\}}\mo$ is
\[
S_{k}(\rho||\sigma)=\begin{cases} 
    0 &  t<|x| \\
     \frac{1}{k-1}\log\frac{c_0\left({\{K_i\},\{K'_j\}}\right)c_0\left({\{K_i\},\{K'_j\}}\right)}{c_0\left({\{K_i\},\{K_i\}}\right)c_0\left({\{K'_j\},\{K'_j\}}\right)}& t>|x|,
\end{cases}\nn\label{s3fulltime3.3}
\]
with  $c_0\left({\{K_i\},\{K'_j\}}\right)$ as the coefficient of the holomorphic two-point correlation function in the sense of \eqref{holo_antiholo}. Eq. \eqref{s3fulltime3.3} shows that RRE can serve as a ``measure" of the distinguishability of two descendant states. Interestingly, even though the states forming the two density matrices $\rho$ and $\sigma$ include both holomorphic and antiholomorphic parts, their RRE ultimately reflects only the information of the holomorphic part. We will discuss the underlying physical picture of this in detail in subsection \ref{subsection4.1}. Additionally, note that in RCFTs, the result of the RRE ultimately depends only on a series of coefficients of two-point functions, which allows us to explore the relationship between RRE and some known metrics in the following subsection.
\subsection{$k^{\rm th}$ relative R\'enyi entropy and trace square distance}
Although we propose that RRE can serve as a measure to distinguish between different states, it is not a suitable measure since it is not symmetric. However, from the results above, such as \eqref{fulltime3.3}, there exist many cases whose RRE are symmetric, arising from the fact that the R\'enyi entropy $\tr \rho^k$ of many states belonging to the same conformal family are universal~\cite{Chen:2015usa}, and the term of the form $\tr \rho\log\sigma$ only depends on the several coefficients of two-point functions \eqref{late3.3} where each one is asymmetrical individually, but when combined, they become symmetrical. Therefore, in this subsection, we will explore the relationship between RRE and the trace square distance \cite{Zhang:2019kwu} under the premise of symmetry.

The trace square distance (TSD) between two reduced-density matrices is given by
\[
\mathbf{T}^{(2)}(\rho_A,\sigma_A):=\frac{\tr \left|\rho_A-\sigma_A\right|^2}{\tr\rho_{vacuum}^2}=\frac{\tr\rho_A^2+\tr\sigma_A^2-2\tr\rho_A\sigma_A}{\tr\rho_{vacuum}^2},
\]
where the factor $\rho_{vacuum}^2$, in particular, removes any UV divergences and allows one to directly express the TSD in terms of four-point functions on the two-sheeted surface $\Sigma_2$ and two-point functions on $\Sigma_1$,
\[
\mathbf{T}^{(2)}(\rho_A,\sigma_A)\equiv&\frac{\la V_\alpha(w_1,\bar w_1)V_\alpha(w_2,\bar w_2)V_\alpha(w_3,\bar w_3)V_\alpha(w_4,\bar w_4)\ra_{\Sigma_2}}{\la V_\alpha(w_1,\bar w_1)V_\alpha(w_2,\bar w_2)\ra^2_{\Sigma_1}}\nn\\
+&\frac{\la V_\beta(w_1,\bar w_1)V_\beta(w_2,\bar w_2)V_\beta(w_3,\bar w_3)V_\beta(w_4,\bar w_4)\ra_{\Sigma_2}}{\la V_\beta(w_1,\bar w_1)V_\beta(w_2,\bar w_2)\ra^2_{\Sigma_1}}\nn\\
-&2\frac{\la V_\alpha(w_1,\bar w_1)V_\alpha(w_2,\bar w_2)V_\beta(w_3,\bar w_3)V_\beta(w_4,\bar w_4)\ra_{\Sigma_2}}{\la V_\alpha(w_1,\bar w_1)V_\alpha(w_2,\bar w_2)\ra_{\Sigma_1}\la V_\beta(w_1,\bar w_1)V_\beta(w_2,\bar w_2)\ra^2_{\Sigma_1}}\label{TSD}
\]
where $V_\alpha$ and $V_\beta$ are operators constructing the two density matrices $\rho$ and $\sigma$ separately. From the above calculations, we can easily obtain the time evolution of TSD under local quenches \eqref{quench3.3},
\[
\mathbf{T}^{(2)}(\rho_A,\sigma_A)=\begin{cases} 
    0 &  t<|x| \\
     \log d_{\alpha}+\log d_{\beta}-2\exp\{(k-1)S_{k}(\rho_A||\sigma_A)\}& t>|x|.
\end{cases}\label{TSDfulltime}
\]
Eq.~\eqref{TSDfulltime} shows that by calculating the trace square distance (TSD), we can obtain the $k^{\rm th}$ RRE. This holds for many cases, as discussed in subsection \ref{subsection3.3}, and requires only the computation of correlation functions on a 2-sheeted Riemann surface. However, the examples we have considered thus far do not cover all possibilities. The conditions the reduced density matrices must satisfy for RRE to exhibit symmetry, and the relationship between RRE and other traditional metrics, remain open questions.

\section{Relative R\'enyi entropy for linear combinations of operators}\label{section4}
In the previous section, we performed a detailed study of the RRE between different states under the excitation of a single operator (either primary or descendant). Given the completeness of the Hilbert space, any state can be represented as a linear combination of a complete set of basis states. Thus, if each state is generated by excitations of a local operator, a composite state can be constructed through a linear combination of these operators' excitations. In this section, we explore the time evolution of the RRE for states generated by linear combinations of operators.

The two density matrices $\rho$ and $\sigma$ we investigate in this section take the form
\[
\rho=\frac{e^{-iHt}V_\alpha(x,-\epsilon)|\Omega\ra\la\Omega|V_\alpha(x,\epsilon)e^{iHt}}{\la V_\alpha(x,\epsilon)V_\alpha(x,-\epsilon)\ra},\quad \sigma=\frac{e^{-iHt}V_\beta(x,-\epsilon)|\Omega\ra\la\Omega|V_\beta(x,\epsilon)e^{iHt}}{\la V_\beta(x,\epsilon)V_\beta(x,-\epsilon)\ra},\label{quench4.1}
\]
where
\[
V_\alpha(x,-\epsilon)=&\sum_{i=1}^MC_iV_i(x,-\epsilon),~\quad V_i(x,-\epsilon)=L_{-\{K_i\}}\bar{L}_{-\{\bar{K}_i\}}\mo(x,-\epsilon), \quad(C_i\in \mathbb{R})\nn\\
V_\beta(x,-\epsilon)=&\sum_{j=1}^{M'}C'_jV'_j(x,-\epsilon),\quad V'_j(x,-\epsilon)=L_{-\{K'_j\}}\bar{L}_{-\{\bar{K}'_j\}}\mo(x,-\epsilon), \quad(C'_j\in \mathbb{R})\label{linear_combination_operator}
\]
and all operators in \eqref{linear_combination_operator} are normalized according to the scheme discussed in \cite{He:2023syy} to avoid adding operators with different dimensions. 

The $k^{\rm th}$ RRE of $V_\alpha$ and $V_\beta$ is
\[
S_{k}(\rho_A||\sigma_A)=\frac{1}{k-1}&\left\{\log\frac{\la V_\alpha(w_1,\bar{w}_1)V_\alpha(w_2,\bar{w}_2)V_\alpha(w_3,\bar{w}_3)\dots V_\alpha(w_{2k},\bar{w}_{2k}\ra_{\Sigma_k}}{\la V_\alpha(w_1,\bar{w}_1)V_\alpha(w_2,\bar{w}_2) \ra_{\Sigma_1}^{k}}\right.\nn\\
&-\left.\log\frac{\la V_\alpha(w_1,\bar{w}_1)V_\alpha(w_2,\bar{w}_2)V_\beta(w_3,\bar{w}_3)\dots V_\beta(w_{2k},\bar{w}_{2k}\ra_{\Sigma_k}}{\la V_\alpha(w_1,\bar{w}_1)V_\alpha(w_2,\bar{w}_2)\ra_{\Sigma_1} \la V_\beta(w_1,\bar{w}_1)V_\beta(w_2,\bar{w}_2) \ra_{\Sigma_1}^{k-1}}\right\},\label{relative_entropy_linear}
\]
which can be evaluated as 
\[
&S_{k}(\rho_A||\sigma_A)=\frac{1}{k-1}\left\{\log\frac{\la V_\alpha(w_1,\bar{w}_1)V_\alpha(w_2,\bar{w}_2)\ra_{\Sigma_k}\dots \la V_\alpha(w_{2k-1},\bar{w}_{2k-1})V_\alpha(w_{2k},\bar{w}_{2k})\ra_{\Sigma_k}}{\la V_\alpha(w_1,\bar{w}_1)V_\alpha(w_2,\bar{w}_2) \ra_{\Sigma_1}^{k}}\right.\nn\\
&-\left.\log\frac{\la V_\alpha(w_1,\bar{w}_1)V_\alpha(w_2,\bar{w}_2)\ra_{\Sigma_k}\la V_\beta(w_3,\bar{w}_3)V_\beta(w_4,\bar{w}_4)\ra_{\Sigma_k}\dots \la V_\beta(w_{2k-1},\bar{w}_{2k-1} )V_\beta(w_{2k},\bar{w}_{2k})\ra_{\Sigma_k}}{\la V_\alpha(w_1,\bar{w}_1)V_\alpha(w_2,\bar{w}_2)\ra_{\Sigma_1} \la V_\beta(w_1,\bar{w}_1)V_\beta(w_2,\bar{w}_2) \ra_{\Sigma_1}^{k-1}}\right\}\nn\\
&=\frac{1}{k-1}(\log 1- \log 1)=0
\]
at early times according to \eqref{early}, \eqref{3.3.2} and \eqref{two_point_function_descendant_early} while its late-time behavior is more complicated. 
Eq.~\eqref{late} imply that for $t>\left|x\right|$, the dominant channel for the holomorphic sector of the $2k$-point function on $\Sigma_1$ is 
\[
(z_1, z_4)\dots(z_{2j+1}, z_{2j+4})\dots(z_2, z_{2k-1})(z_{2k-3}, z_{2k}),
\]
while the dominant channel for the anti-holomorphic sector of the $2k$-point function on $\Sigma_1$ is
\[
(\bar z_1, \bar z_2)(\bar z_3, \bar z_4)\dots(\bar z_{2k-1}, \bar z_{2k}).
\]
Therefore, at late times, the $k^{\rm th}$ relative R\'enyi  entropy of $V_\alpha$ and $V_\beta$ \eqref{relative_entropy_linear} can be evaluated as
\[
&S_{k}(\rho_A||\sigma_A)=\nn\\
&\frac{1}{k-1}\log\left(\frac{\sum^{M}\limits_{i_1,i_2,...,i_{2k}=1} \prod\limits_{u=1}^{k}\Tilde{C}_{i_{2u-1}}\Tilde{C}_{i_{2u}}c_0(\{K_{i_{2u-1}}\},\{K_{i_{2u+2}}\})\bar c_0(\{\bar K_{i_{2u-1}}\},\{\bar{K}_{i_{2u}}\})}{\Big(\sum\limits_{i}\sum\limits_{j}\Tilde{C}_i \Tilde{C}_j c_0(\{K_i\},\{K_j\})\bar c_0(\{\bar K_i\},\{\bar{K}_j\})\Big)^k}\right)\nn\\
-&\frac{1}{k-1}\log \left( \frac{\sum^{M}\limits_{i_1,i_2=1}\sum^{M'}\limits_{j_3,j_4,...,j_{2k}=1}\Tilde{C}_{i_1}\Tilde{C}'_{j_4}C_{i_2}C'_{j_{2k-1}}c_0(\{K_{i_{1}}\},\{K'_{j_{4}}\})c_0(\{K_{i_{2}}\},\{K'_{j_{2k-2}}\})\ {\Xi}}{\Big(\sum\limits_{i,i=1}^M \Tilde{C}_i \Tilde{C}_j c_0(\{K_i\},\{K_j\})\bar c_0(\{\bar K_i\},\{\bar{K}_j\})\Big)\Big(\sum\limits_{p,q=1}^{M'}C'_k C'_l c_0(\{K'_p\},\{K'_q\})\bar c_0(\{\bar K'_p\},\{\bar{K}'_q\})\Big)^{k-1}}\right),\label{late_linear}
\]
where $\Xi\equiv\prod\limits_{u=2}^{k-1}\Tilde{C}'_{j_{2u-1}}\Tilde{C}'_{j_{2u+2}}c_0(\{K'_{i_{2u-1}}\},\{K'_{i_{2u+2}}\})\prod\limits_{v=1}^{k}\bar c_0(\{\bar K_{i_{2v-1}}\},\{\bar{K}_{i_{2v}}\})$ and $\Tilde{C}_i=(-1)^{|K_i|}C_i$, $\Tilde{C}'_i=(-1)^{|K'_i|}C'_i$. The possible positive or negative signs before each coefficient are because $w_1-w_2=-(\bar w_1-\bar w_2)=-2i\epsilon$. 

By introducing several finite-dimensional matrices, $X_{M\times M}$, $\bar X_{M\times M}$, $Y_{M'\times M'}$, $\bar Y_{M'\times M'}$, $\Tilde{R}_{M\times M'}$ and $R^{T}_{M'\times M}$, whose elements take the form 
\[
&X_{ij}={\Tilde{C}_i \Tilde{C}_j} c_0(\{K_{i}\},\{K_{j}\}), \quad \bar X_{ij}=\bar c_0(\{\bar K_{i}\},\{\bar K_{j}\})\nn\\
&Y_{ij}={\Tilde{C}'_i \Tilde{C}'_j} c_0(\{K'_{i}\},\{K'_{j}\}), \quad \bar Y_{ij}= \bar c_0(\{\bar K'_{i}\},\{\bar K'_{j}\})\nn\\
&\Tilde{R}_{ij}=\Tilde{C}_i \Tilde{C}'_j c_0(\{K_{i}\},\{K'_{j}\}),\quad R^{T}_{ij}=\Tilde{C}_j \Tilde{C}'_i c_0(\{K'_{i}\}, \{K_{j}\} )\label{matrixelemet}
\]
Eq.~\eqref{late_linear} can be simplified to
\[
S_{k}(\rho_A||\sigma_A)=\frac{1}{k-1}\log\left(\frac{\tr[(X\bar X^{T})^k]}{[\tr(X\bar X^{T})]^k}\right)-\frac{1}{k-1}\log\left(\frac{\tr[\Tilde{R}\bar Y^{T}(Y\bar Y^{T})^{k-2}R^{T}X^{T}]}{\tr(X\bar X^{T})[\tr(Y\bar Y^{T})]^{k-1}}\right).\label{linear_matrix_form}
\]
As a specific example, we would calculate the RRE for $\partial\mo+\bar\partial \mo$ and $\bar\partial\mo$ in detail. The coefficient for the two-point function of the primary operator $\mo$ has been normalized to $1$, i.e.
\[
\la\mo(w_1,\bar w_1)\mo(w_2,\bar w_2)\ra=\frac{1}{|w_1-w_2|^{4\Delta}}.
\]
Through some simple calculations and by paying particular attention to the positive and negative signs in front of the coefficients, we obtain
\[
X_{2\times 2}=\begin{pmatrix}
     -2\Delta(2\Delta+1) &2\Delta\\
    -2\Delta &1
\end{pmatrix},
\quad\bar X_{2\times 2}=\begin{pmatrix}
     1&2\Delta\\
    -2\Delta &-2\Delta(2\Delta+1)
\end{pmatrix}\nn\\
\Tilde{R}_{2\times 1}=\begin{pmatrix}
    2\Delta\\
    1
\end{pmatrix},\quad R^{T}_{1\times 2}=\begin{pmatrix}
    2 \Delta&1
\end{pmatrix},\quad Y=1,\quad\bar Y=-2\Delta(2\Delta+1).\label{matrix_example}
\]
Replacing all the matrices in \eqref{linear_matrix_form} with \eqref{matrix_example}, the RRE for $\partial\mo+\bar\partial \mo$ and $\bar\partial\mo$ is
\[
S_{k}(\rho^{\partial \mo+\bar \partial \mo}_A||\sigma^{\bar \partial \mo}_A)=\begin{cases} 
    0 &  t<|x| \\
     \frac{k-2}{1-k}\log 2& t>|x|.
\end{cases}
\]

We obtain the RRE for linear combination operators by performing operations on certain finite-dimensional matrices composed of the superposition coefficients and the coefficients of holomorphic and antiholomorphic two-point functions, as shown in Eq.~\eqref{linear_matrix_form}. Specifically, the matrices \( X \) and \( Y \) contain information solely about the holomorphic parts of \( V_\alpha \) and \( V_\beta \), respectively, while the matrices \( \bar{X} \) and \( \bar{Y} \) contain only the antiholomorphic parts of \( V_\alpha \) and \( V_\beta \). Notably, the matrices \( \Tilde{R} \) and \( R^{T} \) capture mixed information from the holomorphic parts of \( V_\alpha \) and \( V_\beta \). However, in the results of \eqref{linear_matrix_form}, we do not require a matrix \( \bar{R} \) representing the mixed antiholomorphic parts of \( V_\alpha \) and \( V_\beta \). This omission is closely related to the quasi-particle picture, which we will discuss further in the next subsection.

\subsection{Relative R\'enyi entropy and quasi-particle}\label{subsection4.1}
As discussed previously, for general descendant operators, RRE depends only on the holomorphic part of the two-point function \eqref{fulltime3.3}, while for linear combination operators, some anti-holomorphic information between the two combined operators is lost, as shown in \eqref{linear_matrix_form}. In this subsection, we will demonstrate that, although relative entropy may initially seem capable of acting as a ``measure" to distinguish between two operators that are indistinguishable in entanglement entropy, this distinguishability is subject to significant limitations which can be explained through the quasi-particle picture.

%We will first explore the relationship between the time evolution of relative R\'enyi entropy and quasi-particle propagation through some simple examples.
Using the results in subsection \ref{subsection3.3}, we can derive the full-time evolution of RRE between some simple reduced density matrices, which are locally excited by $\partial \mo$, $\mo$, $\bar \partial \mo$, and $\bar \partial^2\mo$ separately, i.e.
\[
\rho^1=\frac{\partial\mo|\Omega\ra\la\Omega|\partial\mo}{\la\partial \mo \partial \mo\ra},\quad \sigma^1=\frac{\mo|\Omega\ra\la\Omega|\mo}{\la \mo \mo\ra},\quad \sigma^2=\frac{\bar \partial\mo|\Omega\ra\la\Omega|\bar \partial\mo}{\la \bar \partial\mo \bar \partial\mo\ra},\quad\sigma^3=\frac{\bar \partial^2\mo|\Omega\ra\la\Omega|\bar \partial^2\mo}{\la \bar \partial^2\mo \bar \partial^2\mo\ra},\label{densitymatrices_1}
\]
and their results are
\[
S_{k}(\rho^{1}_A||\sigma^{1}_A)=S_{k}(\rho^{1}_A||\sigma^{2}_A)=S_{k}(\rho^{3}_A||\sigma^{1}_A)=\begin{cases} 
    0 &  t<|x| \\
     \frac{1}{k-1}\log\frac{2\Delta}{2\Delta+1}& t>|x|,
\end{cases}\label{holo4.1.1}
\]
\[
S_{k}(\sigma^{1}_A||\sigma^{2}_A)=S_{k}(\sigma^{1}_A||\sigma^{3}_A)=S_{k}(\sigma^{2}_A||\sigma^{3}_A)=0.\label{holo4.1.2}
\]
We continue to choose the subsystem as $ [0, +\infty)$, with the excitation point initially located on the left side of the subsystem. Eqs.~\eqref{holo4.1.1} and \eqref{holo4.1.2} reveal that although the operators forming the density matrices $ \sigma_1$, $ \sigma_2 $, and $ \sigma_3 $ differ, their reduced density matrices evolve identically over time. From the perspective of relative entropy, a relative entropy of zero implies that the two reduced density matrices are identical. However, does this imply they are indeed identical? Referring to \eqref{densitymatrices_1}, it seems unconvincing to conclude that these density matrices are indeed the same. One possible explanation is that relative entropy may not be reliable for determining whether two reduced-density matrices are distinct.

When the position of the operator excitation is within subsystem A, we find 
\[
&S_{k}(\sigma^{1}_A||\sigma^{2}_A)=\begin{cases} 
    0 &  t<|x| \\
     \frac{1}{k-1}\log\frac{2\Delta}{2\Delta+1}& t>|x|,
\end{cases}
S_{k}(\sigma^{1}_A||\sigma^{3}_A)=\begin{cases} 
    0 &  t<|x| \\
     \frac{1}{k-1}\log\frac{2\Delta+2}{2\Delta+3}& t>|x|,
\end{cases}\nn\\
&S_{k}(\sigma^{2}_A||\sigma^{3}_A)=\begin{cases} 
    0 &  t<|x| \\
     \frac{1}{k-1}\log\frac{\Delta(2\Delta+1)}{(\Delta+1)(2\Delta+3)}& t>|x|.
\end{cases}
\]
Unlike in Eq.~\eqref{holo4.1.2}, changing the excitation position of the operator now makes the reduced density matrices, which were previously indistinguishable by relative entropy, distinguishable. Therefore, relative entropy can still be a helpful tool for determining whether two reduced-density matrices are identical. However, this criterion does not ensure the detectability of all density matrices under any excitation, as it depends on the relative position between the excitation point and the subsystem.

If we consider the reduced density matrix $\rho_A^1$ in Eq.~\eqref{holo4.1.1} as a more general density matrix—such as $\rho_A^2$, excited by a linear combination operator, or $\rho_A^3$, representing a mixed state—can the RRE distinguish between $\sigma_1$, $\sigma_2$, and $\sigma_3$?

For simplicity, let's set 
\[
\rho^2=\frac{(\partial\mo+\bar\partial\mo)|\Omega\ra\la\Omega|(\partial\mo+\bar\partial\mo)}{\la(\partial\mo+\bar\partial\mo)(\partial\mo+\bar\partial\mo)\ra},\ \mathrm{and}\ \rho^3=\frac{1}{2}\frac{\partial\mo|\Omega\ra\la\Omega|\partial\mo}{\la\partial\mo\partial\mo\ra}+\frac{1}{2}\frac{\bar\partial\mo|\Omega\ra\la\Omega|\bar\partial\mo}{\la\bar\partial\mo\bar\partial\mo\ra},
\]
and it is easy to derive the following relation
\[
S_{k}(\rho^{2}_A||\sigma^{1}_A)=S_{k}(\rho^{2}_A||\sigma^{2}_A)=S_{k}(\rho^{2}_A||\sigma^{3}_A)=\begin{cases} 
    0 &  t<|x| \\
     \frac{k-2}{1-k}\log2& t>|x|,
\end{cases}\nn\\
S_{k}(\rho^{3}_A||\sigma^{1}_A)=S_{k}(\rho^{3}_A||\sigma^{2}_A)=S_{k}(\rho^{3}_A||\sigma^{3}_A)=\begin{cases} 
    0 &  t<|x| \\
    S_{k}(\rho^{3}_A)-\frac{1}{k-1}\log \frac{\Delta}{2\Delta+1} & t>|x|,
\end{cases}\label{linear_and_mixed_results}
\]
where $S_{k}(\rho^{3}_A)$ represents the $k^{\rm th}$ R\'enyi entropy of $\rho^3_A$. Eq.~\eqref{linear_and_mixed_results} shows that, regardless of whether the reference state is a linear combination state or a mixed state, $\sigma_1$, $\sigma_2$, and $\sigma_3$ remain indistinguishable\footnote{If we consider an unknown density matrix $*$ and only have access to its RRE with respect to $\rho_{2}$ or $\rho_{3}$ (as given in Eq.~\eqref{linear_and_mixed_results}), this information alone is insufficient to distinguish whether $*$ corresponds to $\sigma_{1}$, $\sigma_{2}$, or $\sigma_{3}$.
}. The quasi-particle picture explains why relative entropy may sometimes fail to distinguish between density matrices that appear entirely different.

\begin{figure}[htbp]
  \centering
  \includegraphics[width=1\linewidth]{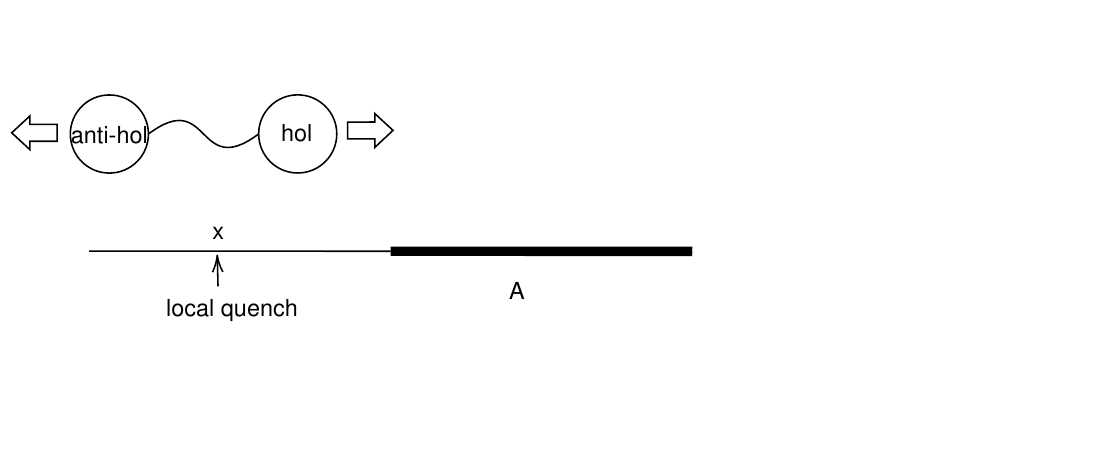}
  \caption{The quasi-particle moving to the left contains only the anti-holomorphic information of the operator, while the quasi-particle moving to the right contains only the holomorphic information of the operator.}\label{figquench1}
\end{figure}

The insertion of a non-chiral operator excites both left- and right-moving quasi-particles, while a purely holomorphic operator excites only a right-moving quasi-particle, and a strictly anti-holomorphic operator excites only a left-moving quasi-particle, as illustrated in Fig.~\ref{figquench1}.

\begin{figure}[htbp]
  \includegraphics[width=1\linewidth]{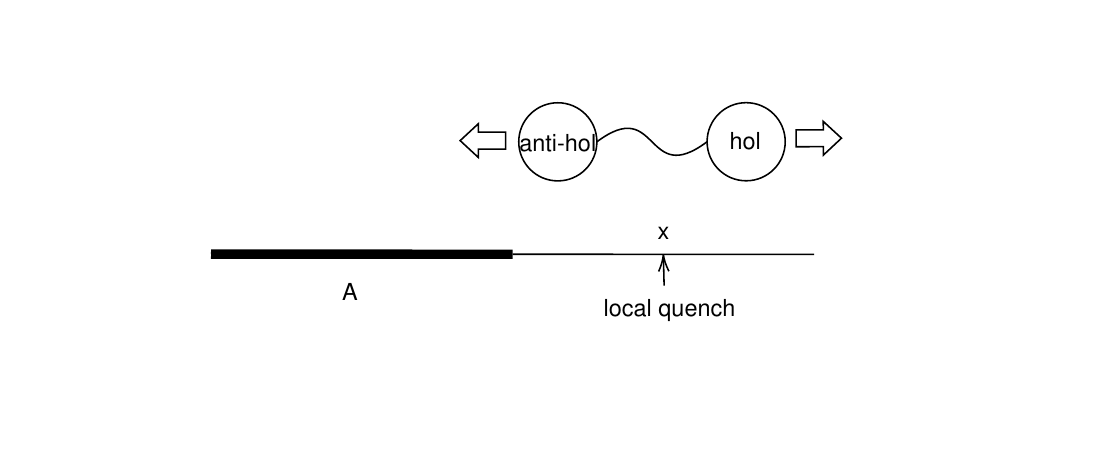}
  \caption{We select subsystem \(A\) as \((-\infty, 0]\), and the excitation point of the operator is on the right side of \(A\).}\label{figquench2}
\end{figure}

Although the operators that constitute $\sigma^1$, $\sigma^2$, and $\sigma^3$ are distinct, after integrating out the degrees of freedom of $\bar A$, the reduced density matrices $\sigma^1_A$, $\sigma^2_A$, and $\sigma^3_A$ retain only the right-moving (holomorphic) information of the operators, which is identical. This creates the appearance that relative entropy fails to distinguish between them. If we initially select the subsystem as $(-\infty, 0]$ and the operator excitation occurs on the positive side of the $x$-axis, as shown in Fig.~\ref{figquench2}, it is straightforward to verify that
\[
S_{k}(\sigma^{3}_A||\sigma^{1}_A)= S_{k}(\sigma^{3}_A||\sigma^{2}_A)= S_{k}(\sigma^{2}_A||\sigma^{3}_A), \quad 0<t<|x|,\label{Abarearly}
\]
since the left-moving particles have not yet entered subsystem \(A\), which can be regarded a vacuum, while
\[
S_{k}(\sigma^{3}_A||\sigma^{1}_A)\neq S_{k}(\sigma^{3}_A||\sigma^{2}_A)\neq S_{k}(\sigma^{2}_A||\sigma^{3}_A).\quad t>|x|,\label{Abarlate}
\]
since $\sigma^1_A$, $\sigma^2_A$, and $\sigma^3_A$ contain different anti-holomorphic information. { One can easily check that $S_k(\rho^1_A||\sigma^1_A)$ is now always zero when considering its full-time evolution compared with the result in Eq. \eqref{holo4.1.1}, which means $\partial \mo$ and $\mo$ are indistinguishable in this case since they contain the same anti-holomorphic information.}

%However, operators like $\partial \mo$ and $\partial^2 \mo$ become indistinguishable in this case.

Thus, relative (R\'enyi) entropy has limitations in distinguishing reduced density matrices, and these limitations are highly dependent on the position of the operator excitations relative to the location of the subsystem. In other words, rather than asserting that relative entropy fails to distinguish different operators, it is more accurate to state that it can only distinguish information from operators that have entered or exited the subsystem.

\section{Relative R\'enyi entropy and entanglement wedge}\label{section5}
 The physical information contained in a given region \( A \) in CFT can correspond to the information in a specific region \( M_A \) in AdS gravity, known as the entanglement wedge \cite{Czech:2012bh, Wall:2012uf, Headrick:2014cta}. This region is defined as the area enclosed by the subsystem \( A \) and the extremal surface \( \Gamma_A \) \cite{Ryu:2006bv}. For a long time, how entanglement wedges emerge from CFT remained unclear until \cite{Suzuki:2019xdq} discovered a sharp structure that reproduces the expected entanglement wedge for 2D holographic CFTs using the Bures metric\footnote{Recent progress has revealed the relationship between the Bures metric and subregion complexity \cite{Gerbershagen:2024qlz}.}. The Bures metric captures the distinguishability of states with different excitations. Given this, we ask: can relative entropy, another ``measure," also reveal the structure of entanglement wedges? In this section, we will demonstrate that RRE may lead to similar results in determining the geometry of entanglement wedges.
 %Physical information included in a given region $A$ in CFT can correspond to that in a certain region $M_A$ in AdS gravity, and this region is given by the entanglement wedge \cite{Czech:2012bh,Wall:2012uf,Headrick:2014cta}, i.e. the region surronden by the subsysetem $A$ and the extremal surface $\Gamma_A$ \cite{Ryu:2006bv}. For a long time, the emergence of entanglement wedges from CFT itself remained unclear until \cite{Suzuki:2019xdq} identified a sharp structure that reproduces the expected entanglement wedge for 2D holographic CFTs using the Bures metric. The Bures metric measures the distinguishability of states with different excitations, so as another "measure", can relative entropy also provide the structure of entanglement wedges? In this section, we will show that RRE may have similar results deriving the geometry of entanglement wedges.
 %[] found a sharp structure and  pointed out how entanglement wedges emerge from CFT itself. In this section, we will show that, from a quantum information perspective, RRE may have similar results deriving the geometry of entanglement wedges.  

 The density matrix considered in this section is constructed by a locally excited primary operator $\mo(w,\bar w)$ in a 2D CFT on a complex plane $R^2$, where we set $(w,\bar w)=(x+i\tau,x-i\tau)$, i.e.
 \[
 \rho(w,\bar w)=\mo(w,\bar w)|\Omega\ra\la\Omega|\mo^\dagger(\bar w,w).
 \]
 We can neglect its backreaction in the gravity dual when assuming that the conformal dimension $h$ of $\mo$ satisfies $1\ll h\ll c$. This allows us to approximate the two-point function  $\langle \mathcal{O}(w, \bar{w}) \mathcal{O}(\bar{w}, w) \rangle$ using the geodesic length in the gravity dual between the two points  $(w, \bar{w})$  and  $(\bar{w}, w)$  on the boundary $ \eta \to 0$  of Poincaré $\rm AdS_3$, described by the metric:
\[
ds^2 = \eta^{-2}(d\eta^2 + dx^2 + d\tau^2).
\] 
%and can approximate the two-point function $\la \mo(w,\bar w)\mo(\bar w,\bar w)\ra$ by the geodesic length in the gravity dual between two points $ (w,\bar w)$ and  $(\bar w, w)$ on the boundary $\et\to0$ of Poincar\'e $\rm AdS_3$,
In the projection on the bulk time slice \(\tau = 0\), the state \(\rho_A(w, \bar{w})\) is dual to a bulk excitation at the bulk point \((\eta, x) = (\tau, x)\), which is defined by the intersection of the time slice \(\tau = 0\) with the geodesic. According to the entanglement wedge reconstruction, two excited bulk states cannot be distinguished if both excitations are located outside the region \(M_A\). However, they become distinguishable if at least one excitation is inside \(M_A\).
%By projection on the bulk time slice $\tau=0$, the state $\rho_A(w,\bar w)$ is dual to a bulk excitation at a bulk point $(\et,x)=(\tau,x)$ defined by the intersection between the time slice $\tau=0$ and the geodesic. The entanglement wedge reconstruction argues that we cannot distinguish two excited bulk states when both excitations are outside of $M_A$, while we can distinguish them if at least one of them is inside of $M_A$.

Relative entropy is an effective measure of quantum information that distinguishes between two quantum states, specifically density matrices. We choose the subsystem $A$ to be an interval $0\leq x \leq L$ at $\tau=0$ for convenient, and the boundary of $M_A$ in the CFT is given by
\[
|w-L/2|=L/2.\label{ew}
\]
In this section, we use the sandwiched {RRE}\footnote{ As we will see, our adoption of the sandwiched RRE in this section is primarily motivated by consistency with the analytic continuation scheme originally defined for collision relative entropy in \cite{Lashkari:2014yva}. While alternative approaches, such as the Petz-$\alpha$ relative entropy or the RRE used elsewhere in this work, could also generate similar results, they correspond to distinct analytic continuation prescriptions. }
\[
S_{\alpha}(\rho\mid\mid\rho^{\prime})=-\frac{1}{1-\alpha}\ln \tr\left(\rho^{\frac{1-\alpha}{2\alpha}}{\rho}^\prime\rho^{\frac{1-\alpha}{2\alpha}}\right)^{\alpha}\label{sandwiched}
\]
where $\rho$ and $\rho'$ are generated by two different locally excited states, i.e.
\[
\rho=\frac{\mo(w,\bar w)|\Omega\rangle\langle \Omega|\mo^{\dagger}(\bar w, w)}{\langle \mo^{\dagger}(\bar w, w)\mo(w,\bar w)\rangle},\quad \rho'=\frac{\mo(w',\bar w')|\Omega\rangle\langle\Omega|\mo^{\dagger}(\bar w', w')}{\langle \mo^{\dagger}(\bar w', w')\mo(w',\bar w')\rangle},
\]
to reproduce the entanglement wedge.

Eq.~\eqref{sandwiched} can be obtained using the replica method and the conformal map $z^k=\frac{w}{w-L}$ when defining $A_{m,n}$ and taking $n=\alpha$ and $m=\frac{1-\alpha}{2\alpha}$, where
\[
A_{m,n}=&\tr[(\rho^m\rho'\rho^m)^n]\nn\\
=&\prod_{i=1}^{2k}\left|k^{-1}(z_i)^{1-k}\right|^{2h}\cdot \prod_{j=1}^k |(z_{2j-1})^k-(z_{2j})^k|^{4h} 
 \langle \mo  ^\dagger(z_1)\mo  (z_2)... \mo  ^\dagger(z_{2k-1})\mo(z_{2k})\rangle_{\Sigma_1}\cdot
\frac{Z^{(k)}}{(Z^{(1)})^k},\label{eq5.1}
\]
and 
\[
&z_1=\left(\frac{-x-i\tau}{L-x-i\tau}\right)^{1/k},\quad z_2=\left(\frac{-x+i\tau}{L-x+i\tau}\right)^{1/k}\nn\\
&z_{2s+1}=e^{(2\pi i/k)s}z_1,\quad z_{2s+2}=e^{(2\pi i/k)s}z_2,\quad(s=1,2,\dots,k-1;k=(2m+1)n).
\]
The entanglement wedge geometry is available only when considering holographic CFTs, so we use the generalized free field approximation to evaluate $A_{n,m}$ in holographic CFTs. When $w$ and $w'$ are outside of the entanglement wedge \eqref{ew} or equally $|z_{2j-1}-z_{2j}|<|z_{2j-2}-z_{2j-1}|$, and choosing $w\approx w'$, the 2$k-$point function on $\Sigma_1$ is approximated as \cite{Kusuki:2019hcg}
\[
 \langle \mo^\dagger(z_1)\mo  (z_2)... \mo  ^\dagger(z_{2k-1})\mo(z_{2k})\rangle_{\Sigma_1}
\approx   \prod_{j=1}^k \langle  \mo^\dagger(z_{2j-1})\mo(z_{2j})\rangle_{\Sigma_1} =  \prod_{j=1}^k |z_{2j-1}-z_{2j}|^{-4h},        
\]
and this leads to $A_{n,m}\approx 1$ and $ S_{\alpha}(\rho\mid\mid\rho^{\prime})\approx 0$. This agrees with the AdS/CFT expectation that we cannot distinguish two bulk excitations outside the entanglement wedge.

 When $w$ and $w'$ are inside of the entanglement wedge \eqref{ew}, the approximation is
 \begin{equation}
     \begin{split}
 \langle O  ^\dagger(z_1)O  (z_2)... O  ^\dagger(z_{2k-1})O(z_{2k})\rangle
\approx   \prod_{j=1}^k \langle  O^\dagger(z_{2j-2})O(z_{2j-1})\rangle =  \prod_{j=1}^k |z_{2j-2}-z_{2j-1}|^{-4h}.        
     \end{split}
 \end{equation}
and thus
 \begin{equation}
A_{\alpha,\frac{1-\alpha}{2\alpha}}=
\frac{|z'-\bar z'|^{4h\alpha}|z-\bar z|^{4h\alpha}}{|z'-\bar z|^{8h\alpha} } 
=
\frac{|w'-\bar w'|^{4h\alpha}|w-\bar w|^{4h\alpha}}{|w'-\bar w|^{8h\alpha} } .\label{eq5.2}
\end{equation}
When $\alpha=1/2$, \eqref{eq5.2} reproduces the fidelity in \cite{Suzuki:2019xdq}, however, when $\alpha\to1$, $S_{\alpha}(\rho\mid\mid\rho^{\prime})$ is diverge\footnote{When calculating relative entropy, divergence is common. The obstruction to analytic continuation in our work might stem from needing to account for subleading terms in the conformal block.}. 
Surprisingly, if $\alpha=2$, \eqref{eq5.2} would give rise to a finite result, the collision relative entropy \cite{Lashkari:2014yva}
\[
S_{2}(\rho\mid\mid\rho^{\prime})=\ln \frac{|w'-\bar w'|^{8h}|w-\bar w|^{8h}}{|w'-\bar w|^{16h} }.\label{eq5.3}
\] 
We can expand \( w' \) and \( \bar{w}' \) around \( w \) and \( \bar{w} \) as:
\[
w' = w + \delta dw \quad \bar{w}' = \bar{w} + \delta \bar{d}w,
\]
where \( \delta \) is an infinitesimal parameter. By expanding \( S_{2}(\rho\mid\mid\rho^{\prime}) \) around \( \delta = 0 \) and then omitting \( \delta \) in the final expression, we obtain the metric:
\[
ds^2 = \frac{2h}{\tau^2}(dx^2 + d\tau^2),
\] 
{which is the Hellinger metric\cite{Kusuki:2019hcg}.} Similarly, in a 2D holographic CFT with a circle compactification $x\sim x+2\pi$, we can obtain the metric
\[
ds^2=\frac{2h}{\sinh^2\tau}(dx^2+d\tau^2)
\]
if $w$ is inside the wedge.

Therefore, the collision relative entropy can produce a Hellinger metric proportional to the AdS metric on a time slice and precisely reproduce the expected entanglement wedge from quantum information in 2D holographic CFTs.

\section{Conclusions and discussions}\label{section6}
This paper investigates the RRE in local operator quenches of RCFTs and holographic CFTs. In RCFTs, the RRE between vacuum and primary operators diverges, as shown in \eqref{vacuum_primary}, due to limitations of the regularization scheme typically used for entanglement entropy in the vacuum state. For descendant operators within the same conformal family (including primaries as exceptional cases), the RRE often exhibits monotonic time evolution, as illustrated in \eqref{fulltime3.3}. This behavior can be interpreted by quasi-particles propagating at the speed of light into the subsystem: the time the RRE reaches zero depends on the distance between the initial excitation point and the subsystem, and the RRE’s magnitude depends on the ordering of the excitation operators. In certain instances, the RRE also displays symmetry under specific operator excitations, prompting an investigation into its relationship with the trace squared distance metric \eqref{TSDfulltime}.

For two linear combination operators, the RRE during time evolution depends solely on certain finite-dimensional matrices \eqref{matrixelemet}, whose dimensions correspond to the number of descendant operators in each combination and whose elements depend on combination coefficients and operator orders. Some examples demonstrate that, despite structural differences in the operators that form distinct density matrices, their RRE can still be zero. The quasi-particle propagation model suggests that relative (R\'enyi) entropy has limitations in distinguishing reduced density matrices, strongly depending on the position of operator excitations relative to the subsystem. Thus, rather than asserting that relative entropy fails to differentiate between operators, it is more accurate to state that it only distinguishes information from operators that have entered or exited the subsystem.

Finally, we reconstruct the geometry of the entanglement wedge from a quantum information perspective. In holographic CFTs, collision relative entropy induces a Hellinger metric, ultimately revealing the sharp structure discussed in \cite{Suzuki:2019xdq}, which provides the geometric structure of the entanglement wedge.

It is important to note that the RRE studied in this paper cannot be analytically continued in either RCFTs or holographic CFTs. {Similar considerations have been explored in previous works, such as~\cite{Caputa:2014vaa}, where the Rényi entanglement entropy in free four-dimensional $U(N)$ Yang-Mills theory was found to scale as
\begin{equation}
\Delta S_A^{(n)} \simeq \frac{Jn - 1}{n - 1} \log 2,\footnote{In \cite{Suzuki:2019xdq}, $J$ is the number of scalar fields in the local operator. }
\end{equation}
which diverges in the von Neumann limit $n \to 1$. The authors of~\cite{Caputa:2014vaa} attribute this divergence to the breakdown of the large-$N$ expansion, suggesting that connected diagrams must be resummed to obtain the correct leading contributions. In our case, one might expect a similar issue since our computation includes only the identity fusion channel. However, we emphasize that in rational CFTs, the identity channel is dominant. This is supported by the fact that the vacuum block alone correctly reproduces the entanglement entropy evolution following local operator quenches in rational CFTs~\cite{He:2014mwa}. We aim to investigate this in future work.}
%The obstruction to analytic continuation in our work might stem from needing to account for subleading terms in the conformal block beyond just the identity operator fusion channel.} 
 Additionally, since the RRE is sometimes symmetric between two density matrices, exploring the sufficient and necessary conditions that lead to this symmetry remains an interesting open question. We would like to leave a further detailed investigation on them for future work.

\subsection*{Acknowledgements}
We thank Rohit Mishra, Masahiro Nozaki, Yuan Sun, Long Zhao, and Yang Liu for valuable discussions on this work.
H. O., H. Z., and Z. Z. are supported by the Science and Technology Development Plan Project of Jilin Province, China Grant No. 20240101326JC.
S. H. acknowledges financial support from the Max Planck Partner Group and the Natural Science Foundation of China, Grants No. 12475053 and No. 12235016. H. O. is also supported by the National Natural Science Foundation of China Grant No. 12205115.

%(No. 12475053, No. 12075101, No. 12235016, No. 12347209).

\appendix

{\section{Useful formulae for computing the k-th RRE}\label{app111}}
{In this section, we provide some useful formulae for computing the $2k$-point functions in the $k$-th RRE \eqref{eq2.2}. Firstly, it is clear that the first term in the first line of \eqref{eq2.2} is just the R\'enyi entropy of $\rho$ and has been investigated explicitly in \cite{He:2014mwa, Chen:2015usa}, while the second term in \eqref{eq2.2} can be evaluated with the help of the usual conformal mapping of $\Sigma_k$ to the complex plane $\Sigma_1$,
\begin{align}
& \la \mo(w_1,\bar w_1)\mo(w_2,\bar w_2)\mo'(w'_3,\bar w'_3)\dots\mo'(w'_{2k},\bar w'_{2k})\ra_{\Sigma_k}\nonumber\\
    =&\prod_{i=1}^2\left(\frac{\partial z_{i}}{\partial w_i}\right)^{\Delta}\left(\frac{\partial \bar z_{i}}{\partial \bar w_i}\right)^{\bar\Delta}\prod_{j=3}^{2k}\left(\frac{\partial z'_{j}}{\partial w'_j}\right)^{\Delta'}\left(\frac{\partial \bar z'_{j}}{\partial \bar w'_j}\right)^{\bar\Delta'}\la \mo(z_1,\bar z_1)\mo(z_2,\bar z_2)\mo'(z'_3,\bar z'_3)\dots\mo'(z'_{2k},\bar z'_{2k})\ra_{\Sigma_1}+...
\end{align}
with
\[
&z^k=w,\quad\quad\quad{\mathrm{if}\quad\ A=[0,\infty)},\nn\\
    &z^k=\frac{w}{w-l},\quad{\mathrm{if}\quad\ A=[0,l]}.\label{conformal_map}
\]
In the above, we take the chiral and antichiral conformal weights of $\mathcal{O}$ to be $\Delta$ and $\bar\Delta$, respectively, with the same assignment for $\mathcal{O}'$. The symbol ``$...$'' denotes additional contributions that appear under the conformal transformation \eqref{conformal_map} when either $\mathcal{O}$ or $\mathcal{O}'$ is a descendant.}

{Upon Wick rotating the Euclidean time $\tau\to it$, for density matrices $\rho$ and $\sigma$ in \eqref{eq2.1}, the complex coordinates on $\Sigma_k$ become
\[
w_{1}&=x+t-i\epsilon,\quad w_{2}=x+t+i\epsilon,\nn\\
\bar{w}_{1}&=x-t+i\epsilon,\quad\bar{w}_{2}=x-t-i\epsilon,\nn\\
w_{2j-1}&=x+t-i\epsilon,\quad w_{2j}=x+t+i\epsilon,\nn\\
\bar{w}_{2j-1}&=x-t+i\epsilon,\quad\bar{w}_{2j}=x-t-i\epsilon,\nn\\
w_{2j-1}'&=x'+t-i\epsilon,\quad w_{2j}'=x'+t+i\epsilon,\nn\\
\bar{w}_{2j-1}'&=x'-t+i\epsilon,\quad\bar{w}_{2j}'=x'-t-i\epsilon,\quad(j=2,3,...,k).
\]
For convenience, unless otherwise specified, we select the subsystem $A$  to be $[0,\infty)$ in our paper, and the 2$k$ points $z_1,z_2,...,z_{2k}$ in the $z$-coordinates are given by
\[
z_{1}=&\text{e}^{\frac{\pi i}{k}}(-x-t+i\e)^{\frac{1}{k}},\quad \bar z_{1}=\text{e}^{-\frac{\pi i}{k}}(-x+t-i\e)^{\frac{1}{k}},\nn\\
z_{2}=&\text{e}^{\frac{\pi i}{k}}(-x-t-i\e)^{\frac{1}{k}},\quad \bar z_{2}=\text{e}^{-\frac{\pi i}{k}}(-x+t+i\e)^{\frac{1}{k}},\nn\\
z_{2j+1}=&\text{e}^{2\pi i\frac{j+1/2}{k}}(-x-t+i\e)^{\frac{1}{k}},\quad \bar z_{2j+1}=\text{e}^{-2\pi i\frac{j+1/2}{k}}(-x+t-i\e)^{\frac{1}{k}},\nn\\
z_{2j+2}=&\text{e}^{2\pi i\frac{j+1/2}{k}}(-x-t-i\e)^{\frac{1}{k}},\quad \bar z_{2j+2}=\text{e}^{-2\pi i\frac{j+1/2}{k}}(-x+t+i\e)^{\frac{1}{k}},\quad (j=1,...,k-1).\,\nn\\
z_{2j+1}'=&\text{e}^{2\pi i\frac{j+1/2}{k}}(-x'-t+i\e)^{\frac{1}{k}},\quad \bar z_{2j+1}'=\text{e}^{-2\pi i\frac{j+1/2}{k}}(-x'+t-i\e)^{\frac{1}{k}},\nn\\
z_{2j+2}'=&\text{e}^{2\pi i\frac{j+1/2}{k}}(-x'-t-i\e)^{\frac{1}{k}},\quad \bar z_{2j+2}'=\text{e}^{-2\pi i\frac{j+1/2}{k}}(-x'+t+i\e)^{\frac{1}{k}},\quad (j=1,...,k-1).\label{A=INFzcoordin}
\]}
{\section{Relative R\'enyi entropies for two locally excited states}}
This section presents the details of the $k^\text{th}$ relative R\'enyi entropy growth calculation for two locally excited states within the same conformal family.

{\subsection{ $|\Omega\rangle$ and $\mo(x)|\Omega\rangle$}\label{appb1ooo}
Two density matrices in this case are given by 
\[
\rho=\frac{e^{-iHt}\mo(x,-\epsilon)|\Omega\ra\la\Omega|\mo(x,\epsilon)e^{iHt}}{\la\mo(x,\epsilon)\mo(x,-\epsilon)\ra},\quad \quad\sigma=\frac{|\Omega\ra\la\Omega|}{\la\Omega|\Omega\ra}.
\]}
{The corresponding $k^{\rm th}$ RRE reads
\[
S_{k}(\rho||\sigma)=\frac{1}{k-1}\log \frac{\la\mo(w_1,\bar w_1)\dots\mo(w_{2k},\bar w_{2k})\ra_{\Sigma_k}}{\la\mo(w_1,\bar w_1)\mo(w_2,\bar w_1)\ra_{\Sigma_1}^k}-\frac{1}{k-1}\log\frac{\la \mo(w_1,\bar w_1)\mo(w_2,\bar w_2)\mathbbm{1}\dots\mathbbm{1}\ra_{\Sigma_k}}{\la\mo(w_1,\bar w_1)\mo(w_2,\bar w_1)\ra_{\Sigma_1}}.\label{app3.1.1}
%=&\frac{(2\epsilon)^{-4\Delta}}{(2\epsilon)^{-2\Delta}(2k\sin \frac{\pi}{k}(t+x))^{-2\Delta}}
\]}
{The second term on the RHS of \eqref{app3.1.1}, by utilizing the conformal map \eqref{conformal_map}, can be reformulated as 
\[
\tr(\rho \sigma^{k-1})=\frac{\left|\frac{d w_1}{d z_1}\right|^{-2\Delta}\left|\frac{d w_2}{d z_2}\right|^{-2\Delta}\la \mo(z_1,\bar z_1)\mo(z_2,\bar z_2)\mathbbm{1}\dots\mathbbm{1}\ra_{\Sigma_1}}{\la\mo(w_1,\bar w_1)\mo(w_2,\bar w_1)\ra_{\Sigma_1}}\label{app3.1.2},
\]
At early times, we had
\[
z_{2i-1}-z_{2i}\approx &e^{2\pi i \frac{i-\frac{1}{2}}{k}} \frac{2 i \epsilon}{k}(-x-t)^{\frac{1-k}{k}}\approx0,\nn\\
\bar z_{2i-1}-\bar z_{2i}\approx &e^{-2\pi i \frac{i-\frac{1}{2}}{k}} \frac{-2 i \epsilon}{k}(-x+t)^{\frac{1-k}{k}}\approx0.\label{early}
\]
Therefore \eqref{app3.1.2} is
\[
\tr(\rho \sigma^{k-1})=\frac{k^{-4\Delta}(x^2-t^2)^{\frac{-2(k-1)\Delta}{k}}k^{4\Delta}(x^2-t^2)^{\frac{2(k-1)\Delta}{k}}\left|2\epsilon\right|^{-4\Delta}}{\left|2\epsilon\right|^{-4\Delta}}=0.
\]
So, the RRE of the primary and identity operators was zero at early times.}

{Next, we would discuss the late-time behavior of \eqref{app3.1.2}. The numerator of the logarithm function can be evaluated as
\[
&\left|\frac{d w_1}{d z_1}\right|^{-2\Delta}\left|\frac{d w_2}{d z_2}\right|^{-2\Delta}\la \mo(z_1,\bar z_1)\mo(z_2,\bar z_2)\mathbbm{1}\dots\mathbbm{1}\ra_{\Sigma_1}\nn\\
=&(k z_1^{k-1} k z_2^{k-1}k \bar z_1^{k-1} k \bar z_2^{k-1})^{-\Delta}(z_1-z_2)^{-2\Delta}(\bar z_1-\bar z_2)^{-2\Delta}=\left[2\epsilon 2k\sin \frac{\pi }{k}(t+x)\right]^{-2\Delta}.
\]
Therefore, at late times, the $k^{\rm th}$ RRE of primary operator and identity operator saturates to
\[
S_{k}(\rho||\sigma)=\frac{1}{k-1}\log\frac{(2\epsilon)^{-2\Delta}}{\left(2k\sin \frac{\pi }{k}(t+x) \right)^{-2\Delta}}-\log d_{\mo}.\label{3.1.3}
\]
From \eqref{app3.1.2} and \eqref{3.1.3}, we derive the full-time evolution of the RRE for the vacuum state, and the primary state is 
\[
S_{k}(\rho||\sigma)=\begin{cases} 
    0 &  t<|x| \\
    \frac{2\Delta}{k-1}\log \frac{k\sin \frac{\pi }{k}(t+x)}{\epsilon}-\log d_{\mo} & t>|x|.
\end{cases}\label{vacuum_primary}
\]}
%As time evolves, the $k^{\rm th}$ RRE of the primary operator and identity operator diverges when we take the regulator $\epsilon\to 0$.  The reason for the divergence of RRE is that the regularization we take in \eqref{eq2.1} is no longer valid since the vacuum state does not have high-energy modes to suppress. %\textcolor{red}{In other words, $\sigma$ constructed from the vacuum state is not regularized while $\rho$ is regularized, so when we consider RRE with this local quench, it always diverges. } 

{\subsection{$L_{-n}\mathcal{O}(x)|\Omega\rangle$ and $\mathcal{O}(x)|\Omega\rangle$}\label{appsb1}}
The two density matrices are constructed from a primary operator and a descendant operator belonging to the same conformal family, i.e.
\[
\rho=\frac{e^{-iHt}\mo(x,-\epsilon)|\Omega\ra\la\Omega|\mo(x,\epsilon)e^{iHt}}{\la\mo(x,\epsilon)\mo(x,-\epsilon)\ra},\quad \sigma=\frac{e^{-iHt}L_{-n}\mo(x,-\epsilon)|\Omega\ra\la\Omega|L_{-n}\mo(x,\epsilon)e^{iHt}}{\la L_{-n}\mo(x,\epsilon)L_{-n}\mo(x,-\epsilon)\ra}.
\]

In terms of \cite{DiFrancesco:1997nk}, the two-point function of $L_{-n}\mo$ and $L_{-n}\mo$ on $\Sigma_1$ reads
\[
&\la L_{-n}\mo(w_1,\bar{w}_1)L_{-n}\mo(w_2,\bar{w}_2)\ra_{\Sigma_1}\nn\\
=&\frac{1}{12} (-1)^n (w_1-w_2)^{-2n}\frac{1}{|w_{12}|^{4\D}} \nn\\
&\left(\frac{\Gamma (2n) \left(c n^2 \left(n^2-1\right)^2  +24 \Delta  (2n) (2n+1) ( n^2-1)\right)}{\Gamma (n+2) \Gamma (n+2)}+12 \Delta  (\Delta   (n+1)^2+2)\right).\label{3.2.2}
\]
Here, $c$ is the central charge. As shown in \cite{Chen:2015usa,He:2023eap}, the $2k$-point function on $\Sigma_k$ can be evaluated as
\[
&\la\mo(w_1,\bar{w}_1)\mo(w_2,\bar{w}_2)L_{-n}\mo(w_3,\bar{w}_3)\dots
 L_{-n}\mo(w_{2k},\bar{w}_{2k})\ra_{\Sigma_k}\nn\\
\sim&\mathcal{F}(w_1,w_2,\dots,w_{2k},n,\Delta)\la\mo(z_1,\bar{z}_1)\mo(z_2,\bar{z}_2)L_{-n}\mo(z_3,\bar{z}_3)\dots
 L_{-n}\mo(z_{2k},\bar{z}_{2k})\ra_{\Sigma_1}
+\dots\label{3.2.3}
\]
where 
\[
\mathcal{F}(w_1,w_2,\dots,w_{2k},n,\Delta)=\big(\prod_{i=1}^{2k}|w_i'|^{-2\D}\big)(w'_3)^{-n}\dots(w'_{2k-1})^{-n}(w'_{2k})^{-n}
\]
is the leading factor coming from the conformal transformation between correlation functions on $\Sigma_k$ and correlation functions on $\Sigma_1$, and the ellipsis in \eqref{3.2.3} denotes terms contributing to lower-order singularity in the correlation functions. 

According to \eqref{early}, the $2k$-point correlation function on $\Sigma_1$ would factorize to
\[
&\la\mo(z_1,\bar{z}_1)\mo(z_2,\bar{z}_2)L_{-n}\mo(z_3,\bar{z}_3)\dots
 L_{-n}\mo(z_{2k},\bar{z}_{2k})\ra_{\Sigma_1}\nn\\
=&\la\mo(z_1,\bar{z}_1)\mo(z_2,\bar{z}_2)\ra_{\Sigma_1}\la L_{-n}\mo(z_3,\bar{z}_3)L_{-n}\mo(z_4,\bar{z}_4)\ra_{\Sigma_1}\dots
 \la L_{-n}\mo(z_{2k-1},\bar{z}_{2k-1})L_{-n}\mo(z_{2k},\bar{z}_{2k})\ra_{\Sigma_1}.\label{3.2.4}
\]
From \eqref{3.2.2} and \eqref{3.2.3} and replacing all the coordinates with \eqref{A=INFzcoordin}, at early times, we find the 
two-point functions on $\Sigma_k$ and on $\Sigma_1$ are the same, i.e.
\[
\la L_{-n}\mo(w_{2i-1},\bar w_{2i-1})L_{-n}\mo(w_{2i},\bar w_{2i})\ra_{\Sigma_k}=\la L_{-n}\mo(w_{2i-1},\bar w_{2i-1})L_{-n}\mo(w_{2i},\bar w_{2i})\ra_{\Sigma_1},\quad (i=1,2,\dots).\label{twopointfunction_early}
\]
Therefore, combining \eqref{early}, \eqref{3.2.1}, \eqref{3.2.2} and \eqref{twopointfunction_early}, the $k^{\rm th}$ RRE of the primary operator and descendant operator at early times is 
\[
S_{k}(\rho||\sigma)=\frac{1}{k-1}(\log 1-\log 1)=0.\label{early3.2}
\]
Based on $\eqref{A=INFzcoordin}$, it can be found that when $t>\left|x\right|$, the $2k$ holomorphic coordinates and the $2k$ anti-holomorphic coordinates approach each other in distinct pairings \cite{Guo:2022sfl} 
\[
&z_{2i-1}-z_{2(i+1)}\approx \frac{-2i\epsilon}{k(x+t)} z_{2i-1}\approx 0,\nn\\
&\bar z_{2i-1}-\bar z_{2i}\approx  \frac{2i\epsilon}{k(x-t)}\bar z_{2i-1}\approx 0.\label{late}
\]
The $2k$-point correlation function on $\Sigma_1$ in \eqref{3.2.3} now can be evaluated as 
\[
&\la\mo(z_1,\bar{z}_1)\mo(z_2,\bar{z}_2)L_{-n}\mo(z_3,\bar{z}_3)\dots
 L_{-n}\mo(z_{2k},\bar{z}_{2k})\ra_{\Sigma_1}\nn\\
=&\la\mo(z_1)L_{-n}\mo(z_4)\dots L_{-n}\mo(z_{2j+1})
 L_{-n}\mo(z_{2j+4})\dots L_{-n}\mo(z_{2k-1})
 \mo(z_{2})\ra_{\Sigma_1}\nn\\
\times&\la\mo(\bar{z}_1)\mo(\bar{z}_2)\mo(\bar{z}_3)\dots
 \mo(\bar{z}_{2k})\ra_{\Sigma_1}
\nn\\
\sim&(F_{00}[\mo])^{k-1}\la\mo(z_1) L_{-n}\mo(z_4)\ra_{\Sigma_1}\dots\la L_{-n}\mo(z_{2j+1}) L_{-n}\mo(z_{2j+4})\ra_{\Sigma_1}\dots\la L_{-n}\mo(z_{2k-1})\mo(z_2)\ra_{\Sigma_1}\nn\\
&\times\la\mo(\bar z_1)\mo(\bar z_2)\ra_{\Sigma_1}\dots\la\mo(\bar z_{2k-3})\mo(\bar z_{2k-2})\ra_{\Sigma_1}\la\mo(\bar z_{2k-1})\mo(\bar z_{2k})\ra_{\Sigma_1}
\]
where we formally decompose the operator $\mo(z,\bar z)$ into a product of a holomorphic operator $\mo(z)$ and an anti-holomorphic operator $\mo(\bar z)$, in the sense of its multi-points correlation function \[&\la\mo(w_1,\bar w_1)\dots\mo(w_k,\bar w_k)\ra\nn\\
=&f(w_1,w_2,\dots,w_k) \bar f(\bar w_1,\bar w_2,\dots,\bar w_k)\nn\\
=&\la\mo(w_1 )\dots\mo(w_k)\ra\la\mo(\bar w_1)\dots\mo(\bar w_k)\ra,\label{holo_antiholo}\]
where $f\ \text{and}\ \bar f$ are the holomorphic part and anti-holomorphic part of the correlation function separately, and we often pick up the proper channel to expand the 2$k$-point function into the holomorphic and the anti-holomorphic part, giving rise to the factor $(F_{00}[\mo])^{k-1}$\footnote{$F_{00}[\mo]$ coincides with the inverse of the quantity called quantum dimension $d_\mo$:
$F_{00}[\mo] = \frac{1}{d_\mo}=\frac{S_{00}}{S_{0\mo}}$,
where $S_{ab}$ is the modular $S$ matrix of the RCFT.}.

Changing back into the $w$-coordinate, with the leading divergent term being transformed homogeneously and keeping the most divergent term, we can find
\[
&\la\mo(w_1,\bar{w}_1)\mo(w_2,\bar{w}_2)L_{-n}\mo(w_3,\bar{w}_3)\dots
 L_{-n}\mo(w_{2k},\bar{w}_{2k})\ra_{\Sigma_k}\nn\\
%\sim&(F_{00}[\mo])^{k-1}\mathcal{F}(w_1,w_2,\dots,w_{2k},n,\Delta)\nn\\
%&\la\mo(z_1) L_{-n}\mo(z_4)\ra_{\Sigma_1}\dots\la L_{-n}\mo(z_{2j+1}) L_{-n}\mo(z_{2j+4})\ra_{\Sigma_1}\dots\la L_{-n}\mo(z_{2k-1})\mo(z_2)\ra_{\Sigma_1}\nn\\
%\times&\la\mo(\bar z_1)\mo(\bar z_2)\ra_{\Sigma_1}\dots\la\mo(\bar z_{2k-3})\mo^{\dagger}(\bar z_{2k-2})\ra_{\Sigma_1}\la\mo(\bar z_{2k-1})\mo^{\dagger}(\bar z_{2k})\ra_{\Sigma_1}\nn\\
\sim&(F_{00}[\mo])^{k-1}\la\mo(z_1) L_{-n}\mo(z_4)\ra_{\Sigma_k}\dots\la L_{-n}\mo(z_{2j+1}) L_{-n}\mo(z_{2j+4})\ra_{\Sigma_k}\dots\la L_{-n}\mo(z_{2k-1})\mo(z_2)\ra_{\Sigma_k}\nn\\
\times&\la\mo(\bar z_1)\mo(\bar z_2)\ra_{\Sigma_k}\dots\la\mo(\bar z_{2k-3})\mo(\bar z_{2k-2})\ra_{\Sigma_k}\la\mo(\bar z_{2k-1})\mo(\bar z_{2k})\ra_{\Sigma_k}.\label{3.2.5}
\]
As pointed out in \cite{He:2023eap}, at the late time, the holomorphic part of the two-point function for two operators belonging to the same conformal family on $\Sigma_k$ and $\Sigma_1$ has the following relation:
\[
&\la L_{-n}\mo(w_{2j+1}) L_{-n}\mo(w_{2j+4})\ra_{\Sigma_k}\nn\\
\sim&(k z_{2j+1}^{k-1})^{-\Delta-n}(k z_{2j+4}^{k-1})^{-\Delta-n}\frac{C_0(n,n)}{(z_{2j+1}-z_{2j+4})^{2\Delta+2n}}\nn\\
\sim&e^{-2\pi i(1+j)(2\Delta+2n)}\la L_{-n}\mo(w_{1}) L_{-n}\mo(w_{2})\ra_{\Sigma_1}.\label{3.2.6}
\]
where we introduce $C_0 (n,n)$  as the coefficient of two-point function on the $\Sigma_1$. The relation of the anti-holomorphic part of the two-point function for two operators belonging to the same conformal family on $\Sigma_k$ and $\Sigma_1$ can be derived similarly,
\[
\la \bar{L}_{-n}\mo(\bar w_{2j+1})\bar{L}_{-m}\mo(\bar w_{2j+2})\ra_{\Sigma_k}\sim e^{2\pi i(1+j)(2\Delta+m+n)}\la\bar{L}_{-n}\mo(\bar w_{1})\bar{L}_{-m}\mo(\bar w_{2})\ra_{\Sigma_1}.\label{3.2.7}
\]
Utilizing \eqref{3.2.1}, \eqref{3.2.5}, \eqref{3.2.6}, and \eqref{3.2.6} and combining with the early time result \eqref{early3.2}, we derive the full-time evolution of the RRE for a primary state and a descendant state
\[
S_{k}(\rho||\sigma)=\begin{cases} 
    0 &  t<|x| \\
     \frac{1}{k-1}\log \frac{12(n+1)^2 \Delta^2}{\left(\frac{\Gamma (2n) \left(c n^2 \left(n^2-1\right)^2  +24 \Delta  (2n) (2n+1) ( n^2-1)\right)}{\Gamma (n+2) \Gamma (n+2)}+12 \Delta  (\Delta   (n+1)^2+2)\right)}& t>|x|.
\end{cases} \label{fulltime3.2}
\]
%\textcolor{red}{Where $c$ is the central charge which has not been defined previously, Please check!}{\color{blue} ZX: C has mentioned in Eq.(18)}

{\subsection{Two general descendant states}\label{appa2}}
Let the two density matrices $\rho$ and $\sigma$ take the form of
\[
\rho=&\frac{e^{-iHt}L_{-\{K_i\}}\bar{L}_{-\{\bar{K}_i\}}\mo(x,-\epsilon)|\Omega\ra\la\Omega|L_{-\{K_i\}}\bar{L}_{-\{\bar{K}_i\}}\mo(x,\epsilon)e^{iHt}}{\la L_{-\{K_i\}}\bar{L}_{-\{\bar{K}_i\}}\mo(x,\epsilon)L_{-\{K_i\}}\bar{L}_{-\{\bar{K}_i\}}\mo(x,-\epsilon)\ra},\nn\\
\sigma=&\frac{e^{-iHt}L_{-\{K'_j\}}\bar{L}_{-\{\bar{K}'_j\}}\mo(x,-\epsilon)|\Omega\ra\la\Omega|L_{-\{K'_j\}}\bar{L}_{-\{\bar{K}'_j\}}\mo(x,\epsilon)e^{iHt}}{\la L_{-\{K'_j\}}\bar{L}_{-\{\bar{K}'_j\}}\mo(x,\epsilon)L_{-\{K'_j\}}\bar{L}_{-\{\bar{K}'_j\}}\mo(x,-\epsilon)\ra},\label{quench3.3}
\]
According to the properties \eqref{early}, the early-time behavior of $2k$-point correlation function in \eqref{s33.3.1} can be evaluated as
\[
&\la L_{-\{K_i\}}\bar{L}_{-\{\bar{K}_i\}}\mo(w_1,\bar{w}_1)L_{-\{K_i\}}\bar{L}_{-\{\bar{K}_i\}}\mo(w_2,\bar{w}_2)L_{-\{K'_j\}}\bar{L}_{-\{\bar{K}'_j\}}\mo(w_3,\bar{w}_3)\dots L_{-\{K'_j\}}\bar{L}_{-\{\bar{K}'_j\}}\mo(w_{2k},\bar{w}_{2k})\ra_{\Sigma_k}\nn\\
\sim&\la L_{-\{K_i\}}\bar{L}_{-\{\bar{K}_i\}}\mo(w_1,\bar{w}_1)L_{-\{K_i\}}\bar{L}_{-\{\bar{K}_i\}}\mo(w_2,\bar{w}_2)\ra_{\Sigma_k}\la L_{-\{K'_j\}}\bar{L}_{-\{\bar{K}'_j\}}\mo(w_3,\bar{w}_3)L_{-\{K'_j\}}\bar{L}_{-\{\bar{K}'_j\}}\mo(w_4,\bar{w}_4)\ra_{\Sigma_k}\nn\\
&\dots\la L_{-\{K'_j\}}\bar{L}_{-\{\bar{K}'_j\}}\mo(w_{2j-1},\bar{w}_{2j-1})L_{-\{K'_j\}}\bar{L}_{-\{\bar{K}'_j\}}\mo(w_{2j},\bar{w}_{2j})\ra_{\Sigma_k}\dots\nn\\
&\la L_{-\{K'_j\}}\bar{L}_{-\{\bar{K}'_j\}}\mo(w_{2k-1},\bar{w}_{2k-1})L_{-\{K'_j\}}\bar{L}_{-\{\bar{K}'_j\}}\mo(w_{2k},\bar{w}_{2k})\ra_{\Sigma_k},\label{3.3.2}
\]
where similar to the discussion in \eqref{twopointfunction_early}, for two general descendant operators, we still have
\[
&\la L_{-\{K'_j\}}\bar{L}_{-\{\bar{K}'_j\}}\mo(w_{2i-1},\bar{w}_{2i-1})L_{-\{K'_j\}}\bar{L}_{-\{\bar{K}'_j\}}\mo(w_{2i},\bar{w}_{2i})\ra_{\Sigma_k}\nn\\
\sim&\la L_{-\{K'_j\}}\bar{L}_{-\{\bar{K}'_j\}}\mo(w_{1},\bar{w}_{1})L_{-\{K'_j\}}\bar{L}_{-\{\bar{K}'_j\}}\mo(w_{2},\bar{w}_{2})\ra_{\Sigma_1},\quad(i=1,2,\dots,k).\label{two_point_function_descendant_early}
\]
Therefore, the early-time behavior of $k^{\rm th}$ RRE of $L_{-\{K_i\}}\bar{L}_{-\{\bar{K}_i\}}\mo$ and $L_{-\{K'_j\}}\bar{L}_{-\{\bar{K}'_j\}}\mo$ is
\[
S_{k}(\rho||\sigma)=\frac{1}{k-1}(\log 1-\log 1)=0.\label{early3.3}
\]
Furthermore, according to \eqref{late}, the late-time behavior of $2k$-point correlation function of $L_{-\{K_i\}}\bar{L}_{-\{\bar{K}_i\}}\mo$ and $L_{-\{K'_j\}}\bar{L}_{-\{\bar{K}'_j\}}\mo$ can be evaluated as
\[
&\la L_{-\{K_i\}}\bar{L}_{-\{\bar{K}_i\}}\mo(w_1,\bar{w}_1)L_{-\{K_i\}}\bar{L}_{-\{\bar{K}_i\}}\mo(w_2,\bar{w}_2)L_{-\{K'_j\}}\bar{L}_{-\{\bar{K}'_j\}}\mo(w_3,\bar{w}_3)\dots L_{-\{K'_j\}}\bar{L}_{-\{\bar{K}'_j\}}\mo(w_{2k},\bar{w}_{2k})\ra_{\Sigma_k}\nn\\
\sim&(F_{00}[\mo])^{k-1}\la L_{-\{K_i\}}\mo(w_1)L_{-\{K'_j\}}\mo(w_4)\ra_{\Sigma_k}\dots\la L_{-\{K'_j\}}\mo(w_{2j+1})L_{-\{K'_j\}}\mo(w_{2j+4})\ra_{\Sigma_k}\dots\nn\\
&\la L_{-\{K'_j\}}\mo(w_{2k-3})L_{-\{K'_j\}}\mo(w_{2k})\ra_{\Sigma_k}\la L_{-\{K'_j\}}\mo(w_{2k-1})L_{-\{K_i\}}\mo(w_{2})\ra_{\Sigma_k}\nn\\
\times&\la \bar{L}_{-\{\bar{K}_i\}}\mo(\bar{w}_1)\bar{L}_{-\{\bar{K}_i\}}\mo(\bar{w}_2)\ra_{\Sigma_k}\la \bar{L}_{-\{\bar{K}'_j\}}\mo(\bar{w}_3)\bar{L}_{-\{\bar{K}'_j\}}\mo(\bar{w}_4)\ra_{\Sigma_k}\dots\la \bar{L}_{-\{\bar{K}'_j\}}\mo(\bar{w}_{2j-1})\bar{L}_{-\{\bar{K}'_j\}}\mo(\bar{w}_{2j})\ra_{\Sigma_k}\nn\\
&\dots\la \bar{L}_{-\{\bar{K}'_j\}}\mo(\bar{w}_{2k-1})\bar{L}_{-\{\bar{K}'_j\}}\mo(\bar{w}_{2k})\ra_{\Sigma_k}.
\]
Similar to the discussion in \eqref{3.2.6} and \eqref{3.2.7}, for two general descendant operators, we have
\[
&\la L_{-\{K_i\}}\mo(w_{2l+1}) L_{-\{K'_j\}}\mo(w_{2l+4})\ra_{\Sigma_k}
\sim e^{-2\pi i(1+l)(2\Delta+\left|K_i\right|+|K'_j|)}\la L_{-\{K_i\}}\mo(w_{2l+1}) L_{-\{K'_j\}}\mo(w_{2l+4})\ra_{\Sigma_1},\nn\\
&\la \bar{L}_{-\{\bar{K}_i\}}\mo(\bar w_{2l+1})\bar{L}_{-\{\bar{K}'_j\}}\mo(\bar w_{2l+2})\ra_{\Sigma_k}\sim e^{2\pi i(1+l)(2\Delta+\left|\bar{K}_i\right|+\left|\bar{K}'_j\right|)}\la\bar{L}_{-\{\bar{K}_i\}}\mo(\bar w_{1})\bar{L}_{-\{\bar{K}'_j\}}\mo(\bar w_{2})\ra_{\Sigma_1},\label{two_point_function_descendant_late}
\]
where $|K_i|\equiv\sum_{j=1}^{n_i}k_{ij},~ |\bar K_i|\equiv\sum_{j=1}^{\bar n_i}\bar k_{ij}$.
Therefore, the late-time behavior of $k^{\rm th}$ RRE of $L_{-\{K'_j\}}\bar{L}_{-\{\bar{K}'_j\}}\mo$ is 
\[
S_{k}(\rho||\sigma)=&\frac{1}{k-1}\log\frac{\la L_{-\{K_i\}}\mo(w_1)L_{-\{K'_j\}}\mo(w_2)\ra_{\Sigma_1}\la L_{-\{K'_j\}}\mo(w_1)L_{-\{K_i\}}\mo(w_{2})\ra_{\Sigma_1}}{\la L_{-\{K_i\}}\mo(w_1)L_{-\{K_i\}}\mo(w_2)\ra_{\Sigma_1}\la L_{-\{K'_j\}}\mo(w_1)L_{-\{K'_j\}}\mo(w_2)\ra_{\Sigma_1}}\nn\\
=&\frac{1}{k-1}\log\frac{c_0\left({\{K_i\},\{K'_j\}}\right)c_0\left({\{K'_j\},\{K_i\}}\right)}{c_0\left({\{K_i\},\{K_i\}}\right)c_0\left({\{K'_j\},\{K'_j\}}\right)},\label{late3.3}
\]
where we denote the notation $c_0\left({\{K_i\},\{K'_j\}}\right)$ as the coefficient of the holomorphic two-point correlation function in the sense of \eqref{holo_antiholo}(similarly for the anti-holomorphic part), i.e.
\[
&\la L_{-\{K_i\}}\bar{L}_{-\{\bar{K}_i\}}\mo(w_1,\bar{w}_1)L_{-\{K'_j\}}\bar{L}_{-\{\bar{K}'_j\}}\mo(w_2,\bar{w}_2)\ra_{\Sigma_1}\nn\\
=&\la L_{-\{K_i\}}\mo(w_1)L_{-\{K'_j\}}\mo(w_2)\ra_{\Sigma_1}\la \bar{L}_{-\{\bar{K}_i\}}\mo(\bar w_1)\bar{L}_{-\{\bar{K}'_j\}}\mo(\bar w_2)\ra_{\Sigma_1}\nn\\
=&\frac{c_0(\{K_i\},\{K'_j\})}{(w_1-w_2)^{2\Delta+|K_i|+|K'_j|}}\frac{\bar c_0(\{\bar K_i\},\{\bar{K}'_j\})}{(\bar{w}_1-\bar{w}_2)^{2\Delta+|\bar K_i|+|\bar K'_j|}}
\]
with $|K_i|\equiv\sum_{j=1}^{n_i}k_{ij}~\text{and}~ |\bar K_i|\equiv\sum_{j=1}^{\bar n_i}\bar k_{ij}$. 
The coefficient of the two-point correlation function for generic descendant operators can be evaluated using the algorithm in Appendix \ref{appendixA} and the program in Appendix \ref{appendixB}. We retain the notation $c_0$ and $\bar c_0$ for convenience in the following content.

From \eqref{early3.3} and \eqref{late3.3}, the full time evolution of $k^{\rm th}$ RRE of $L_{-\{K_i\}}\bar{L}_{-\{\bar{K}_i\}}\mo$ and $L_{-\{K'_j\}}\bar{L}_{-\{\bar{K}'_j\}}\mo$ is
\[
S_{k}(\rho||\sigma)=\begin{cases} 
    0 &  t<|x| \\
     \frac{1}{k-1}\log\frac{c_0\left({\{K_i\},\{K'_j\}}\right)c_0\left({\{K_i\},\{K'_j\}}\right)}{c_0\left({\{K_i\},\{K_i\}}\right)c_0\left({\{K'_j\},\{K'_j\}}\right)}& t>|x|.
\end{cases}\label{fulltime3.3}
\]

\section{Correlation function of two descendant operators}\label{appendixA}
Following the standard way \cite{DiFrancesco:1997nk}, in this section, we can compute the two-point function \[\la L_{-n_1}L_{-n_2}\dots L_{-n_i}\mo(w_1,\bar w_1)L_{-m_1}L_{-m_2}\dots L_{-m_j}\mo(w_2,\bar w_2)\ra.\label{descendant_two_point}\]

Firstly, we assume that the coefficient of the two-point correlation function of the primary operator has been normalized to $1$, i.e.
\[
\la \mo(w_1,\bar w_1)\mo(w_2,\bar w_2)\ra=\frac{1}{w_{12}^{2\Delta}\bar w_{12}^{2\bar\Delta}},
\]
and according to \cite{DiFrancesco:1997nk}, the correlation function taking the form
\[
\la L_{-n_1}L_{-n_2}\dots L_{-n_i}\mo(w_1,\bar w_1)\mo(w_2,\bar w_2)\ra\label{descendant2}
\]
can be evaluated as 
\[
\mathcal{L}_{-n_1}\mathcal{L}_{-n_2}\dots \mathcal{L}_{-n_i} \la \mo(w_1,\bar w_1)\mo(w_2,\bar w_2)\ra
\]
where
\[
\mathcal{L}_{-n}=\frac{(n-1)\Delta}{(w_2-w_1)^n}-\frac{\partial_{w_2}}{(w_2-w_1)^{n-1}}.
\]
Next, We will progressively degrade \eqref{descendant_two_point} step by step into the form of \eqref{descendant2},
\[
&\la L_{-n_1}L_{-n_2}\dots L_{-n_i}\mo(w_1,\bar w_1)L_{-m_1}L_{-m_2}\dots L_{-m_j}\mo(w_2,\bar w_2)\ra\nn\\
=&-\frac{1}{2\pi i}\oint_{\mathcal{C}(w_2)}\frac{\dd w}{(w-w_1)^{n_1-1}}\la T(w)L_{-n_2}\dots L_{-n_i}\mo(w_1,\bar w_1)L_{-m_1}L_{-m_2}\dots L_{-m_j}\mo(w_2,\bar w_2)\ra .\label{1time}
%=&-\frac{1}{2\pi i}\oint_{\mathcal{C}(w_2)}\frac{\dd w}{(w-w_1)^{n_1-1}}\la L_{-n_2}\dots L_{-n_i}\mo(w_1,\bar w_1)L_{-m_1}L_{-m_2}\dots L_{-m_j}\mo(w_2,\bar w_2)\ra\nn\\
\]
For simplicity, we introduce a shorthand notation, denoting $L_{-k_1}L_{-k_2}\dots L_{-k_s}\mo$ as $\mo^{(-k_1,-k_2,\dots,-k_s)}$.

Since
\[
&T(w)L_{-m_1}L_{-m_2}\dots L_{-m_j}\mo(w_2,\bar w_2)\nn\\
\sim&\frac{m_1(m_1^2-1)c/12+2 m_1(\sum\limits_{k=2}^{j} m_{k}+\Delta)}{(w-w_2)^{m_1+2}}\mo^{(-m_2,-m_3,\dots,-m_j)}(w_2,\bar w_2)\nn\\
+&\sum\limits_{k=1}^{m_1-1}\frac{(m_1+k)}{(w-w_2)^{k+2}}\mo^{(-(m_1-k),-m_2,-m_3,\dots,-m_j)}(w_2,\bar w_2)\nn\\
+&\frac{(\sum\limits_{k=1}^{j}m_k+\Delta)}{(w-w_2)^2}\mo^{(-m_1,-m_2,-m_3,\dots,-m_j)}(w_2,\bar w_2)\nn\\
+&\frac{\partial_{w_2}}{w-w_2}\mo^{(-m_1,-m_2,-m_3,\dots,-m_j)}(w_2,\bar w_2),
\]
Eq. \eqref{descendant_two_point} will reduce to 
\[
&\la L_{-n_1}L_{-n_2}\dots L_{-n_i}\mo(w_1,\bar w_1)L_{-m_1}L_{-m_2}\dots L_{-m_j}\mo(w_2,\bar w_2)\ra\nn\\
=&-\frac{1}{2\pi i}\oint_{\mathcal{C}(w_2)}\frac{\dd w}{(w-w_1)^{n_1-1}}\nn\\
&[\frac{m_1(m_1^2-1)c/12+2 m_1(\sum\limits_{k=2}^{j} m_{k}+\Delta)}{(w-w_2)^{m_1+2}}\la \mo^{(-n_2,-n_3,\dots,-n_i)}(w_1,\bar w_1)\mo^{(-m_2,-m_3,\dots,-m_j)}(w_2,\bar w_2)\ra\nn\\
+&\sum\limits_{k=1}^{m_1-1}\frac{(m_1+k)}{(w-w_2)^{k+2}}\la \mo^{(-n_2,-n_3,\dots,-n_i)}(w_1,\bar w_1)\mo^{(-(m_1-k),-m_2,-m_3,\dots,-m_j)}(w_2,\bar w_2)\ra\nn\\
+&\frac{(\sum\limits_{k=1}^{j}m_k+\Delta)}{(w-w_2)^2}\la \mo^{(-n_2,-n_3,\dots,-n_i)}(w_1,\bar w_1)\mo^{(-m_1,-m_2,-m_3,\dots,-m_j)}(w_2,\bar w_2)\ra\nn\\
+& \frac{\partial_{w_2}}{w-w_2}\la \mo^{(-n_2,-n_3,\dots,-n_i)}(w_1,\bar w_1)\mo^{(-m_1,-m_2,-m_3,\dots,-m_j)}(w_2,\bar w_2)\ra]\nn\\
=&(-1)^{m_1}\frac{(n_1+m_1-1)!}{(m_1+1)!(n_1-2)!}\frac{m_1(m_1^2-1)c/12+2 m_1(\sum\limits_{k=2}^{j} m_{k}+\Delta)}{(w_2-w_1)^{m_1+n_1}}\nn\\
&\times\la \mo^{(-n_2,-n_3,\dots,-n_i)}(w_1,\bar w_1)\mo^{(-m_2,-m_3,\dots,-m_j)}(w_2,\bar w_2)\ra\nn\\
+&(-1)^{n_1}\sum\limits_{k=1}^{m_1-1}\frac{(n_1+k-1)!}{(k+1)!(n_1-2)!}\frac{(m_1+k)}{(w_1-w_2)^{n_1+k}}\la \mo^{(-n_2,-n_3,\dots,-n_i)}(w_1,\bar w_1)\mo^{(-(m_1-k),-m_2,-m_3,\dots,-m_j)}(w_2,\bar w_2)\ra\nn\\
+&\frac{(n_1-1)(\sum\limits_{k=1}^{j}m_k+\Delta)}{(w_2-w_1)^{n_1}}\la \mo^{(-n_2,-n_3,\dots,-n_i)}(w_1,\bar w_1)\mo^{(-m_1,-m_2,-m_3,\dots,-m_j)}(w_2,\bar w_2)\ra\nn\\
-&\frac{\partial_{w_2}}{(w_2-w_1)^{n_1-1}}\la \mo^{(-n_2,-n_3,\dots,-n_i)}(w_1,\bar w_1)\mo^{(-m_1,-m_2,-m_3,\dots,-m_j)}(w_2,\bar w_2)\ra.\label{order1}
%\la L_{-n_2}\dots L_{-n_i}\mo(w_1,\bar w_1)(\frac{m_1(m_1^2-1)c/12+2 m_1(\sum\limits_{k=2}^{j} m_{k}+\Delta)}{(w-w_2)^{m_1+2}}\mo^{(-m_2,-m_3,\dots,-m_j)}(w_2,\bar w_2)\nn\\
%+\sum\limits_{k=1}^{m_1-1}\frac{(m_1+k)}{(w-w_2)^{k+2}}\mo^{(-(m_1-k),-m_2,-m_3,\dots,-m_j)}(w_2,\bar w_2)\nn\\
%+\frac{(\sum\limits_{k=1}^{j}m_k+\Delta)}{(w-w_2)^2}\mo^{(-m_1,-m_2,-m_3,\dots,-m_j)}(w_2,\bar w_2)\nn\\
%+\frac{\partial_{w_2}}{w-w_2}\mo^{(-m_1,-m_2,-m_3,\dots,-m_j)}(w_2,\bar w_2))\ra\nn\\
\]
In the final expression of \eqref{order1}, there can be at most \(i-1\) Virasoro generators for the first operator. We can use a similar approach in \eqref{1time} several times to reduce the correlation function \eqref{descendant_two_point} to \eqref{descendant2}.

\section{{Mathematica code}}\label{appendixB}
In this section, we present the Mathematica code to evaluate the coefficient of the two-point correlation function for generic descendant operators, as discussed at the end of Section 3.
\begin{verbatim} 
Clear[LLLOOfunction];
LLLOOfunction[
   formula_?((#[[1]][[1]] // ToString) == 
        "L" && (#[[-1]][[0]] // ToString) == 
        "\[CapitalOmega]" && (#[[-2]][[0]] // ToString) == 
        "\[CapitalOmega]" &)] := Module[{class},
   class["step1"] = formula;
   class["n"] = Length[class["step1"]] - 2;
   class["L_function"] = ((#1 - 
          1) \[CapitalDelta])/((-1)^#1 (\[Omega]1 - \[Omega]2)^#1) #2 \
- \!\(
\*SubscriptBox[\(\[PartialD]\), \(\[Omega]2\)]#2\)/((-1)^(#1 - 
        1) (\[Omega]1 - \[Omega]2)^(#1 - 1)) &;
   class["correlation"] = 
    1/((\[Omega]1 - \[Omega]2)^(
     2 \[CapitalDelta]) (\[Omega]bar1 - \[Omega]bar2)^(
     2 \[CapitalDelta]bar));
   class["value"] = class["correlation"];
   Table[Module[{}, 
     class["value"] = 
      class["L_function"][-class["step1"][[class["n"] - i + 1]][[2]], 
       class["value"]]], {i, 1, class["n"]}];
   class["value"]
   ];
   Clear[LLLOLLLOfunction]
LLLOLLLOfunction[formula_?((#[[1]][[1]] // ToString) == "L" &)] := 
  Module[{xn1, xm1, numn, numm},
   
   xn1 = -formula[[1]][[2]];
   xm1 = Module[{}, 
     For[i = 1, i <= Length[#], i++, 
        Module[{}, 
         If[(#[[i]][[1]] // ToString) == "L", numn = i, Break[]]]] &@
      formula; -formula[[numn + 2]][[2]]];
   numm = Length[formula] - numn - 2;
   If[numm == 0,
    back = LLLLsimplify[formula];
    ,
    kk = ToExpression["k" <> ToString[flag["n"]]];
    flag["func"];
    back = (-1)^xm1 (xn1 + xm1 - 1)!/((xm1 + 1)! (xn1 - 2)!) (
       xm1 (xm1^2 - 1) c/12 + 
        2 xm1 (Sum[-formula[[numn + 1 + ks]][[2]], {ks, 2, 
             numm}] + \[CapitalDelta]))/((-1)^(
        xm1 + xn1) (\[Omega]1 - \[Omega]2)^(xm1 + xn1))
        LLssimplify[(Table[
             If[i == 1 || i == numn + 2, Nothing, #[[i]]], {i, 1, 
              Length[#]}] /. List -> NonCommutativeMultiply &@
          formula)] + (-1)^
        xn1 Sum[(xn1 + ks - 1)!/((ks + 1)! (xn1 - 2)!) (xm1 + 
           ks)/(\[Omega]1 - \[Omega]2)^(xn1 + ks)
          LLssimplify[(Table[
               If[i == 1, Nothing, 
                If[i == numn + 2, Subscript[
                 L, -(xm1 - ks)], #[[i]]]], {i, 1, Length[#]}] /. 
              List -> NonCommutativeMultiply &@formula)], {ks, 1, 
         xm1 - 1}] + ((xn1 - 
          1) (Sum[-formula[[numn + 1 + ks]][[2]], {ks, 1, 
            numm}] + \[CapitalDelta]))/((-1)^
        xn1 (\[Omega]1 - \[Omega]2)^xn1)
        LLssimplify[(Table[
             If[i == 1, Nothing, #[[i]]], {i, 1, Length[#]}] /. 
            List -> NonCommutativeMultiply &@formula)] - 
      1/((-1)^(xn1 - 1) (\[Omega]1 - \[Omega]2)^(xn1 - 1))
        pd@(LLssimplify[(Table[
               If[i == 1, Nothing, #[[i]]], {i, 1, Length[#]}] /. 
              List -> NonCommutativeMultiply &@formula)]);
    back = 
     StringReplace[
       ToString[back, 
        InputForm], {"ks" -> \!\(TraditionalForm\`ToString[kk]\)}] // 
      ToExpression;];
   back
   
   ];
   Clear[OLLLOfunction]
OLLLOfunction[
   formula_?((#[[1]][[0]] // ToString) == "\[CapitalOmega]" &)] := 
  Module[{class},
   class["step1"] = 
    formula /. {\[CapitalOmega][\[Omega]1, \[Omega]bar1] ** 
        A_ ** \[CapitalOmega][\[Omega]2, \[Omega]bar2] -> 
       A ** \[CapitalOmega][\[Omega]2, \[Omega]bar2] ** \
\[CapitalOmega][\[Omega]1, \[Omega]bar1]};
   class["n"] = Length[class["step1"]] - 2;
   class["L_function"] = ((#1 - 
          1) \[CapitalDelta])/(\[Omega]1 - \[Omega]2)^#1 #2 - \!\(
\*SubscriptBox[\(\[PartialD]\), \(\[Omega]1\)]#2\)/(\[Omega]1 - \
\[Omega]2)^(#1 - 1) &;
   class["correlation"] = 
    1/((\[Omega]1 - \[Omega]2)^(
     2 \[CapitalDelta]) (\[Omega]bar1 - \[Omega]bar2)^(
     2 \[CapitalDelta]bar));
   class["value"] = class["correlation"];
   Table[Module[{}, 
     class["value"] = 
      class["L_function"][-class["step1"][[class["n"] - i + 1]][[2]], 
       class["value"]]], {i, 1, class["n"]}];
   class["value"]
   ];
   ppd[x_] := Module[{class},
  \!\(
\*SubscriptBox[\(\[PartialD]\), \(\[Omega]2\)]x\)
  ]
  normalfunction[formula_] := 
  formula /. {(-\[Omega]1 + \[Omega]2)^
       A_ (-\[Omega]bar1 + \[Omega]bar2)^B_ -> (-1)^
       A (\[Omega]1 - \[Omega]2)^A (-1)^
       B (\[Omega]bar1 - \[Omega]bar2)^B} /. {(-1)^(
     A___ - 2 \[CapitalDelta] - 2 \[CapitalDelta]bar) -> (-1)^A};
flag = Module[{class},
   class["n"] = 1;
   class["func"] := Module[{},
     class["n"] = class["n"] + 1;
     ];
   class
   ];
   classify[string_] := Module[{class},
  class["len"] = string // Length;
  If[((#[[1]][[1]] // ToString) == "L" &@
      string) && ((#[[-2]][[0]] // ToString) == "\[CapitalOmega]" &@
      string), class["out"] = LLLOOfunction[string]];
  If[((#[[1]][[1]] // ToString) == "L" &@
      string) && ((#[[-2]][[1]] // ToString) == "L" &@string), 
   class["out"] = LLLOLLLOfunction[string]];
  If[((#[[1]][[0]] // ToString) == "\[CapitalOmega]" &@
      string) && ((#[[-2]][[1]] // ToString) == "L" &@string), 
   class["out"] = OLLLOfunction[string]];
  If[((#[[1]][[0]] // ToString) == "\[CapitalOmega]" &@
      string) && ((#[[-2]][[0]] // ToString) == "\[CapitalOmega]" &@
      string), 
   class["out"] = 
    1/((\[Omega]1 - \[Omega]2)^(
     2 \[CapitalDelta]) (\[Omega]bar1 - \[Omega]bar2)^(
     2 \[CapitalDelta]bar))];
  class["out"]
  ]
  block["formula change"] := Module[{},
   change = Module[{class},
      class["formula"] = input["formula"];
      Print[class["formula"]];
      class["formula2"] = 
       class["formula"] // StringReplace[#, {"<" -> "", ">" -> ""}] &;
      class["formula3"] = class["formula2"] // ToExpression;
      class["back"] = class["formula3"];
      outputp["formula"] = class["formula3"];
      class];
   ];




   input["formula"] = 
  "<\!\(\*SubscriptBox[\(L\), \(-n1\)]\)**\[CapitalOmega][\[Omega]1,\
\[Omega]bar1]**\!\(\*SubscriptBox[\(L\), \(-m3\)]\)**\[CapitalOmega][\
\[Omega]2,\[Omega]bar2]>";(*Input the correlation function to be calculated.*)
block["formula change"]
outputp["formula"] // classify;
% /. LLssimplify -> classify;
% /. pd -> ppd // FullSimplify;
% // Refine[#, n1 \[Element] PositiveIntegers] & // FullSimplify

\end{verbatim}
\bibliographystyle{JHEP}
\bibliography{RelativeRef}
\end{document}